\documentclass[onecolumn, a4paper, accepted=2026-09-16]{quantumarticle}

\pdfoutput=1

\usepackage{amsmath, amsfonts, amssymb, amsthm, dsfont}
\usepackage{bm}
\usepackage{hyperref}
\usepackage{physics}
\usepackage{tabularx}
\usepackage{cancel}
\usepackage{tikz}
\usepackage{pifont}
\usetikzlibrary{arrows.meta}
\usepackage{usefulnotations}
\usepackage[numbers, sort&compress]{natbib}
\newcommand{\lbra}[1]{{\vphantom{\ket{#1}}}_{\text{L}}\!\!\bra{#1}}

\newcommand{\rket}[1]{\ket{#1}_{\!\text{R}}}

\begin{document}
	\title{Evidence for Exceptional Points as Topological Defects}

	\author{Chia-Yi Ju}
	\affiliation{Department of Physics, National Sun Yat-sen University, Kaohsiung 804201, Taiwan}
	\affiliation{Center for Theoretical and Computational Physics, National Sun Yat-sen University, Kaohsiung 804201, Taiwan}
	\affiliation{Research Center for Quantum Computing and Quantum Materials, National Sun Yat-sen University, Kaohsiung 804201, Taiwan}
	\affiliation{Physics Division, National Center for Theoretical Sciences, Taipei 106319, Taiwan}
	\email{chiayiju@mail.nsysu.edu.tw}
	\orcid{0000-0001-7038-3375}
	
	\author{Szu-Ming Chen}
	\affiliation{National Science and Technology Museum, Kaohsiung 807044, Taiwan}

	\maketitle

	\begin{abstract}
		Studies have shown that quantum states reside in a Hilbert space bundle. When a quantum system depends on continuous external parameters, these parameters define additional dimensions in the base space of the bundle. While much of the existing literature focuses on eigenstate subbundles, where geometric properties like Berry curvature arise, this work considers the entire Hilbert space bundle. Although the Hilbert space bundle has been found to be locally flat, suggesting that the system's topology may appear trivial, we revisit this assumption. Specifically, we examine how an arbitrary quantum state evolves when transported along closed parameter loops, a phenomenon characterized by holonomy. Our results demonstrate that nontrivial holonomy can emerge in the presence of exceptional points. Consequently, exceptional points naturally manifest as topological defects. Finally, we show that this nontrivial topology manifests in physical, time-dependent evolutions, providing a simple experimental signature to detect exceptional points by comparing state transport along distinct paths.
	\end{abstract}

	\section{Introduction \label{Sec:Introduction}}

		Non-Hermitian quantum systems are known for their fascinating and unexpected phenomena~\cite{Peng2014, Leykam2017, Quijandria2018, Znojil2023a, Znojil2023b, Znojil2024a, Yang2024, Feyisa2025, Feyisa2025a}. Nevertheless, a Hilbert space metric is essential to ensure self-consistency in the theoretical study of non-Hermitian systems~\cite{Bender1998, Mostafazadeh2003, Bender2004, Bender2007, Brody2013, Znojil2020, Znojil2023, Chen2025, Ju2025}. To extend the validity of the theory from pseudo-Hermitian systems~\cite{Mostafazadeh2010} to more general non-Hermitian systems, including EPs, the Hilbert space is naturally generalized to a Hilbert space bundle~\cite{Mostafazadeh2004, Ju2019, Ju2022}. In this framework, the Hilbert space becomes the fiber space, and time serves as the base space, allowing the Hilbert space to evolve over time.

		Interestingly, when the Hamiltonian depends on $n$ independent continuous parameters, these parameters can naturally introduce emergent dimensions in the base space~\cite{Ju2024}. As a result, the base space extends from $\mathds{R}$, representing time, to $\mathds{R} \times \mathcal{M}^n$, where $\mathcal{M}^n$ denotes the $n$-dimensional parameter space manifold. The local curvature of the corresponding Hilbert space bundle vanishes everywhere, indicating that the bundle is locally flat. This raises the question of whether the Hilbert space bundle in which quantum states reside is globally trivial.

		To answer this question, we turn our attention to exceptional points (EPs)~\cite{Kato1976, Heiss2004, Znojil2020a, Znojil2021, Naikoo2023}, which are widely regarded as the most exotic structures in the non-Hermitian regime.

		One of the most remarkable phenomena involves encircling an EP in parameter space. Studies have shown that the eigenvalues of the Hamiltonian can transition between different Riemann sheets~\cite{Szekeres1959, Dembowski2004, Zhong2018, Oezdemir2019, Arkhipov2023, Arkhipov2024, Lai2024, Beniwal2024, Roy2025}, resulting in a domain larger than that of the original parameter space. Many studies suggest that this phenomenon arises from the underlying topology of the eigenstates near an exceptional point~\cite{Kawabata2019, Hu2022, Yang2022, Okuma2023, Jia2023}.

\begin{figure}[h!]
			\begin{center}
				\begin{tabular}{cccccc}
					(a) & & (b) & & (c) &\vspace{-2.5ex}\\
					& \includegraphics[width = 0.25\textwidth]{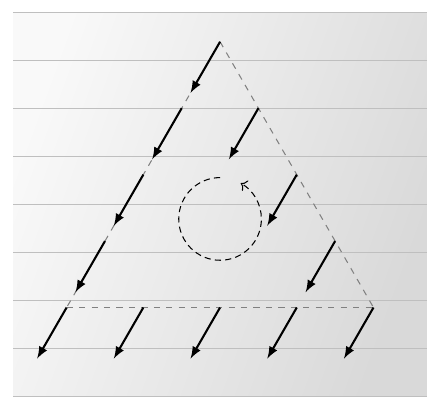} & & \includegraphics[width = 0.25\textwidth]{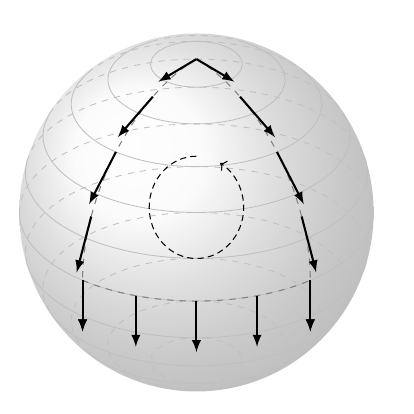} & & \includegraphics[width = 0.27\textwidth]{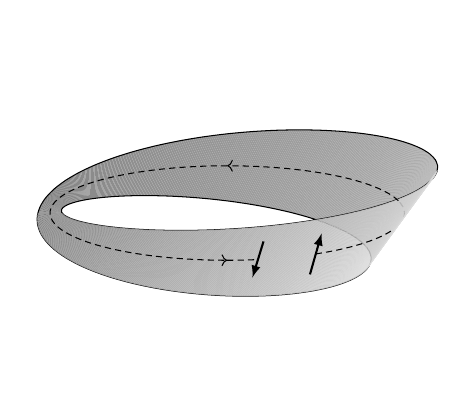}
				\end{tabular}
			\end{center}
			\caption{Parallel transporting a vector around a closed path under different geometries (and topologies). (a) On a flat plane, a vector transported along the dashed path ends up unchanged. (b) On a sphere, a vector at the north pole transported along the dashed path ends up pointing in a different direction from the original. (c) On a flat M\"{o}bius strip, the vector flips its orientation after traveling along the dashed path due to the strip’s nontrivial topology.}
			\label{Fig:CurvedVSFlat}
		\end{figure}

		We therefore study how states change when they evolve along closed paths. In a curved space, a state does not necessarily evolve back to itself; however, in a flat space, we expect the state to return to itself, unless the topology is nontrivial (as illustrated in Fig.~\ref{Fig:CurvedVSFlat}). The relationship between a state evolved along a closed path and its original state is called holonomy. Therefore, a nontrivial holonomy in the Hilbert space bundle suggests that the topological of the system can be nontrivial, since the local curvature is zero.

		From the perspective of the Berry connection and curvature~\cite{Berry1984}, special treatments are required to describe the evolution of states around an EP. For instance, one must consider all relevant eigenstates simultaneously and generalize the Berry connection to a non-Abelian one~\cite{Tang2022, Guo2023}. Alternatively, one can enlarge the domain in which the eigenstates evolve, which leads to the conclusion that the eigenstate swapping phenomenon is geometric rather than topological~\cite{MehriDehnavi2008}. Nevertheless, the Berry connection is a projection of the Hilbert space bundle's connection onto an eigenstate. Similar to the case of a 2-dimensional sphere embedded in a 3-dimensional Euclidean space, where the former has nonzero curvature while the latter is flat{. T}his can lead to some subtle differences in the discussion. Hence, rather than judging from the eigenstates, this study considers the full Hilbert space bundle, which, as previously mentioned, is locally flat; therefore, a trivial holonomy group (the collection of all holonomies) is expected if the topology of the bundle is trivial. However, the nontrivial holonomy of eigenstates transported around a loop encircling an EP strongly suggests that EPs may play an exotic role in the topology of the Hilbert space bundle.

		Therefore, in this study, we demonstrate that in a seemingly simply connected space, the holonomy of a loop around an EP is indeed nontrivial, despite the fact that the local curvature vanishes everywhere. This indicates that the EP acts as a defect in the topological space. Specifically, if the parameter space of the Hamiltonian is denoted by $\mathcal{M}^n$, the full base space of the Hilbert space bundle is given by $\left(\mathds{R} \times \mathcal{M}^n\right) \setminus E$, where $E = \{(t, m) \in \mathds{R} \times \mathcal{M}^n \mid m \text{ is an EP of the Hamiltonian}\}$.

	\section{A Quick Review of the Emergent Dimensions \label{Sec:EmergentDimension}}

		In this section, we provide a brief review on how the states and the geometry depends on parameters (detailed derivations and discussions can be found in \cite{Ju2019, Ju2024}). The main idea is to allow the geometry of the quantum states to be determined by the physical laws rather than being assigned arbitrarily. Consequently, the Hilbert space inner product is relaxed from the conventional $\braket{\Phi(t)}{\Psi(t)}$ to
		\begin{equation}
			\Braket{\Phi(t)}{\Psi(t)} = \bra{\Phi(t)} G(t) \ket{\Psi(t)},
		\end{equation}
		where $t \in \mathds{R}$ denotes time (therefore, the base space is $\mathds{R}$), $\Ket{\Psi(t)} = \ket{\Psi(t)}$, $\Bra{\Phi(t)} = \bra{\Phi(t)} G(t)$, and $G(t)$ is the Hilbert space bundle metric to be determined.

		Nevertheless, for $\Braket{\Phi(t)}{\Psi(t)}$ to qualify as a proper Hilbert space inner product, the metric $G(t)$ has to be Hermitian [$G^\dagger(t) = G(t)$] and positive-definite [$\bra{\chi} G(t) \ket{\chi} \geq 0$ for any $\ket{\chi}$]. To find the governing equation for the metric, we turn our attention to the Sch\"{o}dinger equation in the covariant derivate form, namely,
		\begin{equation}
			\nabla_t \ket{\Psi(t)} \equiv \left[\partial_t + i H(t) \right] \ket{\Psi(t)} = 0, \label{StateTEq}
		\end{equation}
		where $\nabla_t$ is treated as a connection (or a covariant derivative) in the $t$-direction, and $H(t)$ is the Hamiltonian or time-evolution generator. By requiring the metric to be connection-compatible, we obtain
		\begin{align}
			& 0 = \nabla_t \Braket{\Phi(t)}{\Psi(t)} = \nabla_t \bra{\Phi(t)} G(t) \ket{\Psi(t)}\\
			\Rightarrow \hspace{0.5ex} & 0 = \nabla_t G(t) = \partial_t G(t) - i G(t) H(t) + i H^\dagger(t) G(t). \label{MetricTEq}
		\end{align}
		It is worth noting that, from Eq.~\eqref{MetricTEq}, the conventional inner product, i.e., $G(t) = \mathds{1}$, is recovered only when the Hamiltonian is Hermitian.\\
		(On a side note: the metric $G$ is rigorously defined only when the corresponding Hilbert space is finite-dimensional~\cite{Mostafazadeh2010, Ju2019}.)

		\subsection{Single-Parameter Dependent Hamiltonian}

			In the case where the Hamiltonian also depends on a parameter $q$, i.e., $H = H(t, q)$, the metric $G$ should depend on $q$, since it is determined by the Hamiltonian $H(t, q)$. In other words, Eq.~\eqref{MetricTEq} renders
			\begin{equation}
				0 = \partial_t G(t, q) - i G(t, q) H(t, q) + i H^\dagger(t, q) G(t, q). \label{WrongDeriveMetricQEq}
			\end{equation}

			However, the assumption $G = G(t, q)$ is not always valid. Since the base space is no longer $\mathds{R}$, but rather $\mathds{R} \times \mathcal{M}$, where $\mathcal{M}$ is the parameter space, $G$ may not necessarily be a function of $t$ and $q$, but may instead depend on the path taken. In other words, Eq.~\eqref{WrongDeriveMetricQEq} should more accurately be written as
			\begin{equation}
				0 = \partial_t G[\gamma] - i G[\gamma] H(t, q) + i H^\dagger(t, q) G[\gamma], \label{DeriveMetricQEq}
			\end{equation}
			where $G[\gamma]$ emphasizes the path dependence, and $\gamma$ is the corresponding path. For the sake of brevity, we will omit explicit parameter and path dependence when no confusion is expected.

			Utilizing the Hermiticity and positive-definiteness of $G$, it can be deduced that the $q$-derivative of $G$ is formally
			\begin{equation}
				\partial_q G = i G K - i K^\dagger G,
			\end{equation}
			although the operator $K$ remains unknown. By assuming that a quantum state $\ket{\Psi}$ remains physical along any physically allowed path taken, i.e.,
			\begin{align}
				\Braket{\Psi[\gamma]}{\Psi[\gamma]} = 1 \label{Assumption}
			\end{align}
			for any path $\gamma$ in $\mathds{R} \times \mathcal{M}$, we obtain
			\begin{align}
				\partial_q \ket{\Psi} = - i K \ket{\Psi}, \label{DeriveStateQEq}
			\end{align}
			where $K$ can be realized as the $q$-evolution generator, which still remains to be determined.

			As a matter of fact, Eqs.~\eqref{DeriveMetricQEq} and \eqref{DeriveStateQEq}, like the $t$-evolution of the state and the metric, describe parallel transport along the $q$-direction, i.e.,
			\begin{align}
				0 & = \nabla_q G \equiv \partial_q G - i G K + i K^\dagger G. \label{MetricQEq}\\
				0 & = \nabla_q \ket{\Psi} \equiv \left(\partial_q + i K \right) \ket{\Psi}, \label{StateQEq}
			\end{align}

			Combining Eqs.~\eqref{StateTEq} and \eqref{StateQEq}, it follows that $\left[\nabla_t, \nabla_q\right] = 0$, which leads to
			\begin{align}
				0 = \partial_t K - \partial_q H + i \left[H, K\right]. \label{KEq}
			\end{align}
			In other words, the unknown generator $K$ is related to the Hamiltonian $H$ through Eq.~\eqref{KEq}. That is, $K$ can be determined by solving Eq.~\eqref{KEq}.

			Building on the preceding discussion, we are now ready to introduce the curvature two-form, which is defined as
			\begin{align}
				\mathcal{F} = \frac{1}{2}\left(F_{tq} dt \wedge dq + F_{qt} dq \wedge dt\right),
			\end{align}
			where
			\begin{align}
				F_{tq} = - F_{qt} = - i \left[\nabla_t, \nabla_q\right].
			\end{align}
			Since $\left[\nabla_t, \nabla_q\right] = 0$, the local curvature two-form vanishes identically; in other words,
			\begin{align}
				\mathcal{F} = 0. 
			\end{align}
			Therefore, when the full Hilbert space bundle is taken into account, it is locally flat. To avoid potential confusion, we emphasize that the corresponding Berry curvature---defined through the Berry connection, which is the projection of $K$ onto a given eigenstate (i.e., a subbundle)---doesn’t have to be flat.

			Although, $K$ cannot be uniquely determined from Eq.~\eqref{KEq}, different choices of $K$ do not alter the underlying physics (see Appendix~\ref{Appendix:GaugeDOF} for details). To be more specific, it has been shown that the non-uniqueness comes from the gauge degree of freedom in choosing the basis. Therefore, a choice of gauge is required.

			For a time-independent system, i.e., when $\partial_t H = 0$, it is much easier to find $K$ by adopting the gauge condition
			\begin{align}
				[\partial_t K, H] = 0. \label{AdiabaticGauge}
			\end{align}
			Imposing this condition and taking an additional $t$-derivative on both sides of Eq.~\eqref{KEq} yields
			\begin{align}
				0 & = \partial_t^2 K - \partial_t \partial_q H + i \left[\cancel{\partial_t H}, K\right] + i \left[H, \partial_t K\right] = \partial_t^2 K - \partial_q \cancel{\partial_t H} - i \bcancel{\left[\partial_t K, H\right]} = \partial_t^2 K.
			\end{align}
			In other words, this gauge condition ensures that $K$ is at most linear in $t$, i.e.,
			\begin{align}
				K = t \p{K}{1} + \p{K}{0},
			\end{align}
			where $\p{K}{1}$ and $\p{K}{0}$ do not depend on $t$, and reduces Eq.~\eqref{KEq} to a set of algebraic equations, namely,
			\begin{align}
				\left\{\begin{array}{l}
					\displaystyle \left[\p{K}{1}, H\right] = 0\\
					\displaystyle \p{K}{1} + i \left[H, \p{K}{0}\right] = \partial_q H
				\end{array}\right..
			\end{align}
			This means that the gauge in Eq.~\eqref{AdiabaticGauge} reduces the differential equation to a set of algebraic equations.

			Therefore, with Eqs.~\eqref{StateQEq} and \eqref{AdiabaticGauge}, we are equipped to study how quantum states evolve in response to changes in the parameter $q$.

		\subsection{Multi-Parameter Dependent Hamiltonian}

			We now extend the discussion from single-parameter dependence to multi-parameter dependence. Assume that the base space $\mathds{R} \times \mathcal{M}^n$ is described by the parameters $\{q^0 = t, q^1, q^2, \dots, q^n\}$, where $\mathcal{M}^n$ is an $n$-dimensional manifold representing the parameter space and the superscript in $q^\mu$ is not the exponent of $q$ but as an index distinguishing the independent variables. As a result, we obtain
			\begin{align}
				0 & = \nabla_\mu \ket{\Psi} \equiv \left(\partial_\mu + i K_\mu \right) \ket{\Psi}, \label{DeriveStateQEqs}\\
				0 & = \nabla_\mu G \equiv \partial_\mu G - i G K_\mu + i K_\mu^\dagger G,
			\end{align}
			where $\mu \in \{0, 1, 2, \dots, n\}$, $\partial_\mu = \dfrac{\partial}{\partial q^\mu}$, and $K_0 = H$.

			Analogous to the single-parameter case, the local curvature two-form also vanishes everywhere, i.e.,
			\begin{align}
				\mathcal{F} = \frac{1}{2} F_{\mu \nu} dq^\mu \wedge dq^\nu = 0,
			\end{align}
			where
			\begin{align}
				F_{\mu \nu} = - i \left[\nabla_\mu, \nabla_\nu\right] = 0, \label{F}
			\end{align}
			and Einstein summation convention is applied. Therefore, the Hilbert space bundle is, again, locally flat. Moreover, from Eq.~\eqref{F}, we find that the operators $K_\mu$ satisfy the following relations:
			\begin{align}
				0 = \partial_\mu K_\nu - \partial_\nu K_\mu + i \left[K_\mu, K_\nu\right]. \label{KEqs}
			\end{align}

			To simplify the derivation of $K_i$, where $i \in \{1, 2, \dots, n\}$ (we adopt the convention from general relativity: Greek indices run from $0$ to $n$, while Latin indices run from $1$ to $n$), we can generalize the gauge condition shown in Eq.~\eqref{AdiabaticGauge} to
			\begin{align}
				[\partial_0 K_i, H] = [\partial_t K_i, H] = 0, \label{AdiabaticGauges}
			\end{align}
			so that when the Hamiltonian does not depend on $t$,
			\begin{align}
				K_i = t \p{K_i}{1} + \p{K_i}{0}, \label{KAdiabaticGauge}
			\end{align}
			where $\p{K_i}{1}$ and $\p{K_i}{0}$ are both $t$-independent operators.

			From Eq.~\eqref{F}, the relationships between $\p{K_i}{0}$, $\p{K_i}{1}$, and $H$ are
			\begin{align}
				\left\{\begin{array}{l}
					\displaystyle \left[\p{K_i}{1}, H\right] = 0\\
					\displaystyle \p{K_i}{1} + i \left[H, \p{K_i}{0}\right] = \partial_i H(q)\\
					\displaystyle \partial_i \p{K_j}{1} - \partial_j \p{K_i}{1} = i \left[\p{K_j}{0}, \p{K_i}{1}\right] - i \left[\p{K_i}{0}, \p{K_j}{1}\right]\\
					\displaystyle \partial_i \p{K_j}{0} - \partial_j \p{K_i}{0} = i \left[\p{K_j}{0}, \p{K_i}{0}\right]
				\end{array}\right.. \label{KRelation}
			\end{align}

			As a result, just as in the single-parameter case, the gauge conditions in Eq.~\eqref{AdiabaticGauges} transforms the differential equations in Eq.~\eqref{KEqs} into algebraic ones.

	\section{Holonomy from the Evolution Generators}

		Holonomy, roughly speaking, describes how a vector changes after being transported around a closed path in the base space, relative to its initial value. In our case, it captures how the quantum state evolves when the parameters in the Hamiltonian are varied along a path that eventually returns to their original values. To study this, we examine the evolution of general quantum states along closed paths in the combined time-parameter space $\mathds{R} \times \mathcal{M}^n$.

		Unlike the Berry connection, the connections defined in Eqs.~\eqref{DeriveStateQEq} and \eqref{DeriveStateQEqs} are not limited to Hamiltonian eigenstates, but can be applied to arbitrary physical states. We can therefore use them to compute the holonomy associated with the vector bundle. Since the holonomy is defined for the full vector bundle, the transformation of an arbitrary state is determined by the holonomy operator $U$, namely, $\ket{\Psi_\text{f}} = U \ket{\Psi_\text{i}}$. Nevertheless, this holonomy is not generally a pure scalar phase (i.e., $\ket{\Psi_\text{f}} \not\propto \ket{\Psi_\text{i}}$ in general), unlike the Berry phase in the Hamiltonian eigenstate case.

		We begin with a formal discussion of a Hamiltonian that depends on $n$ parameters and time $t$, i.e., $H(q^0, q^1, \dots, q^n)$, where $q^0 = t$. To find the holonomy associated with a closed path, we need to understand how states evolve along a path.

		By parameterizing the path $\gamma: [a, b] \to \mathds{R} \times \mathcal{M}^n$ as
		\begin{align}
			\gamma (s) = \left(q_\gamma^0 (s), q_\gamma^1 (s), \dots, q_\gamma^n (s)\right),
		\end{align}
		where $q_\gamma^\mu (s)$ denotes the value of the parameter $q^\mu$ along the path $\gamma$ at point $s$, the evolution of the state $\ket{\Psi[\gamma]}$ satisfies
		\begin{align}
			\frac{d}{ds} \ket{\Psi[\gamma]} = \frac{dq_\gamma^\mu}{ds} \partial_\mu \ket{\Psi[\gamma]} = - i \frac{dq_\gamma^\mu}{ds} K_\mu \ket{\Psi[\gamma]}.
		\end{align}
		We define the evolution operator $U[\gamma](s; a)$ along the path $\gamma$ by
		\begin{align}
			\ket{\Psi[\gamma](s)} = U[\gamma](s; a) \ket{\Psi[\gamma](a)},
		\end{align}
		where $U[\gamma](a; a) = \mathds{1}$ and satisfies the differential equation
		\begin{align}
			\frac{d}{ds} U[\gamma](s; a) = - i \frac{dq_\gamma^\mu}{ds} K_\mu U[\gamma](s; a). \label{EvolutionEq}
		\end{align}
		A side note: if the path $\gamma$ lies entirely along the $q^0$-direction (i.e., the time direction), then $U[\gamma]$ reduces to the standard time-evolution operator.

		As a consequence, the evolution of the state is completely determined by the evolution operator $U[\gamma]$. Therefore, given a closed path $\gamma$, i.e., $\gamma(b) = \gamma(a)$, the corresponding holonomy is determined by $U[\gamma](b; a)$.

		It is worth noting that the collection of $U[\gamma]$, where $\gamma$ is a closed loop starting at a point $\tilde{q} \in \mathds{R} \times \mathcal{M}^n$, forms a group $\mathrm{Hol}_{\tilde{q}}$, called the holonomy group (often simply denoted as $\mathrm{Hol}$ since holonomy groups based at different points are isomorphic to each other). Additionally, since the holonomy groups under different gauge choices (i.e., with different choices of evolution generators $\{K\}$) are also isomorphic to one another (see Appendix~\ref{Appendix:HolonomyClass}), it is sufficient to study the holonomy under whichever gauge is most convenient.

	\section{Trivial and Nontrivial Holonomies \label{Sec:Holonomy}}

		As discussed in Sec.~\ref{Sec:EmergentDimension}, the Hilbert space bundle extended by the parameters is locally flat. This implies that the holonomy group is trivial unless the vector bundle is nontrivial; that is, unless the underlying base space contains defects. In other words, given a closed path $\gamma: [a, b] \to \mathds{R} \times \mathcal{M}^n$ with $\gamma(b) = \gamma(a)$, we should expect $U[\gamma](b; a) = \mathds{1}$ unless the path encloses some topological defects.

		In this section, we demonstrate that $U[\gamma_{\scriptscriptstyle \text{EP}}](b; a) \neq \mathds{1}$ if the path $\gamma_{\scriptscriptstyle \text{EP}}$ encircles an EP. In other words, the EP behaves as a topological defect.

\begin{figure}[h!]
			\begin{center}
				\begin{tabular}{cccc}
					(a) & & (b) &\vspace{-2.5ex}\\
					& \includegraphics[width = 0.35\textwidth]{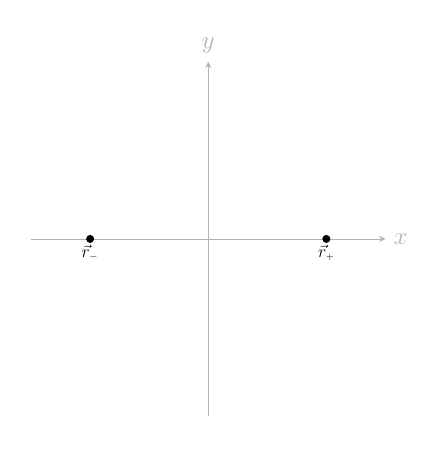} & & \includegraphics[width = 0.35\textwidth]{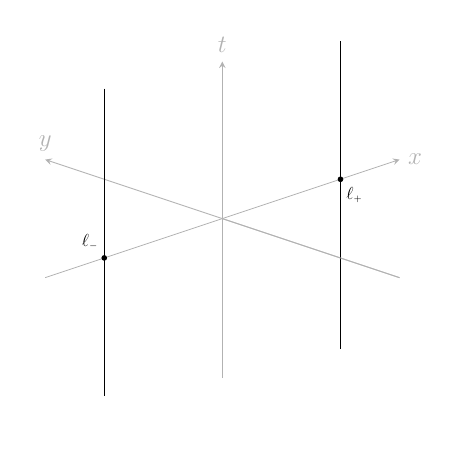}
				\end{tabular}
			\end{center}
			\caption{Illustration of the parameter space and base space of $H(x, y)$. (a) The parameter space $\mathds{R}^2$: Two exceptional points are located in the $x$-$y$ plane at $\vec{r}_\pm = \pm \hat{e}_x$. (b) The base space $\mathds{R} \times \mathds{R}^2$: The two exceptional points become two lines in the base space, shown as the black solid lines. Although they appear as points in the parameter space, when the time dimension is taken into account, they trace out lines $\ell_\pm$ in the base space. The parameter space shown in (a) can be thought of as a specific time slice (e.g., the $t = 0$ slice) of the full base space.}
			\label{Fig:BaseSpace}
		\end{figure}

		Consider a non-Hermitian Hamiltonian of the form
		\begin{align}
			H(x, y) = \begin{pmatrix}
				- i x & 1 + i y\\
				1 + i y & i x
			\end{pmatrix},
		\end{align}
		where $x$ and $y$ are real parameters. Since both $x$ and $y$ can take any real value, the parameter space should therefore be $\mathcal{M}^2 \cong \mathds{R}^2$, which ostensibly suggests that the base space is $\mathds{R} \times \mathds{R}^2 \cong \mathds{R}^3$, which is simply connected. We will soon find that this is, in fact, not the case due to the existence of nontrivial holonomies. Nevertheless, we proceed with this seemingly plausible assumption. In the parameter space $\mathds{R}^2$, there are two EPs located at $\vec{r}_\pm = \pm \hat{e}_x$. In the full base space, they sweep out two lines $\ell_\pm: \mathds{R} \to \mathds{R} \times \mathds{R}^2$, defined as $\ell_\pm = \{(t, \vec{r}) \in \mathds{R} \times \mathds{R}^2 \mid \vec{r} = \vec{r}_\pm\}$, which can be parameterized as $\ell_\pm (t) = \left(t, (\pm 1, 0)\right)$ (see Fig.~\ref{Fig:BaseSpace}). As we will soon see, this leads to a contradiction; therefore, the base space should be $\mathds{R} \times \mathds{R}^2 \setminus \left(\ell_- \cup \ell_+\right)$ rather than all of $\mathds{R} \times \mathds{R}^2$. To simplify notation, we write $(t, (x, y)) \in \mathds{R} \times \mathds{R}^2$ as $(t, x, y)$ provided no ambiguity arises.

		To determine the evolution operator $U[\gamma]$, we need to find the evolution generators in the $x$- and $y$-directions, i.e., $K_x$ and $K_y$ in
		\begin{align}
			0 & = \nabla_x \ket{\Psi} = \left(\partial_x + i K_x\right) \ket{\Psi},\\
			0 & = \nabla_y \ket{\Psi} = \left(\partial_y + i K_y\right) \ket{\Psi}.
		\end{align}
		Hence, we impose the gauge conditions shown in Eq.~\eqref{AdiabaticGauges}, apply the determining equations shown in Eqs.~\eqref{KAdiabaticGauge} and \eqref{KRelation}, and obtain
		\begin{align}
			K_x & = \frac{-1}{\left(1 + i y\right)^2 - x^2} \begin{pmatrix}
				- i x^2 t & \left(1 + i y\right) x t - \dfrac{1 + i y}{2}\\
				\left(1 + i y\right) x t + \dfrac{1 + i y}{2} & i x^2 t
			\end{pmatrix},\\
			K_y & = \frac{1}{\left(1 + i y\right)^2 - x^2} \begin{pmatrix}
				\left(1 + i y\right) x t & i \left(1 + i y\right)^2 t - \dfrac{i x}{2}\\
				i \left(1 + i y\right)^2 t + \dfrac{i x}{2}\ & - \left(1 + i y\right) x t
			\end{pmatrix}.
		\end{align}

		Notice that there are two singularities in both $K_x$ and $K_y$, located at the two EPs $\vec{r}_\pm$ in the parameter space $\mathds{R}^2$. These singularities are, in fact, a feature of EPs, specifically, the evolution generators are generally singular at the EPs~\cite{Tzeng2021, Tu2022, Tu2023, Ju2025a}. This suggests that something nontrivial occurs at the EPs.

		\subsection{Holonomy Without Enclosing an EP}

			\begin{figure}[h!]
				\begin{center}
					\begin{tabular}{cccc}
						(a) & & (b) &\vspace{-2.5ex}\\
						& \includegraphics[width = 0.35\textwidth]{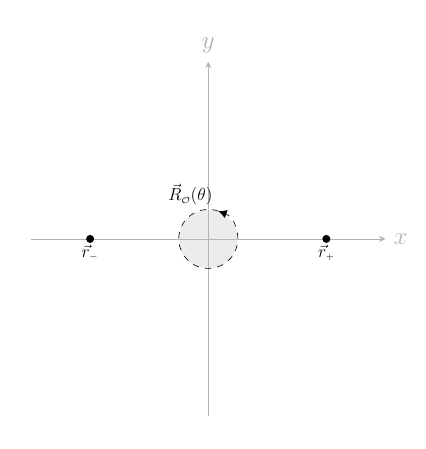} & & \includegraphics[width = 0.35\textwidth]{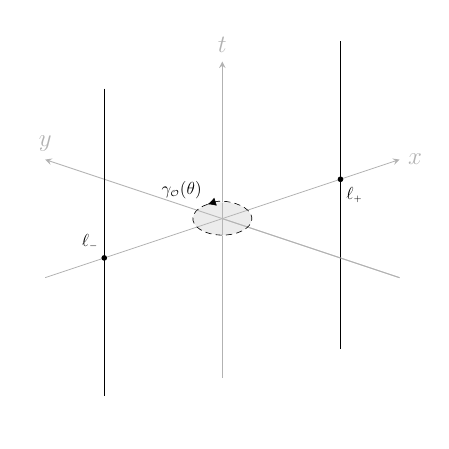}
					\end{tabular}
				\end{center}
				\caption{Transport the state along the path $\gamma_{\scriptscriptstyle \mathcal{O}}$. (a) State evolution in the parameter space $\mathds{R}^2$ along ${\vec{R}_{\scriptscriptstyle \mathcal{O}} (\theta) = r \cos \theta ~ \hat{e}_x + r \sin \theta ~ \hat{e}_y}$, which is the projection of $\gamma_{\scriptscriptstyle \mathcal{O}}$ onto the parameter space. No EPs are encircled by the closed path. (b) State evolution in the base space $\mathds{R} \times \mathds{R}^2$. The path does not enclose either of the lines swept out by the EPs.}
				\label{Fig:EncircleNonEP}
			\end{figure}

			As a demonstration, we first discuss the case in which the state evolves around a region such that no EP is enclosed by the closed path $\gamma_{\scriptscriptstyle \mathcal{O}}: [0, 2 \pi] \to \mathds{R} \times \mathds{R}^2$ (see Fig.~\ref{Fig:EncircleNonEP}). To keep the discussion concrete and simple, we choose the path to be
			\begin{align}
				\gamma_{\scriptscriptstyle \mathcal{O}} (\theta) = (0, x_{\scriptscriptstyle \mathcal{O}}(\theta), y_{\scriptscriptstyle \mathcal{O}}(\theta)) = (0, r \cos \theta, r \sin \theta),
			\end{align}
			where $r$ is a fixed value satisfying $0 < r < 1$, chosen so that the loop does not enclose any EPs. From Eq.~\eqref{EvolutionEq}, we find that
			\begin{align}
				\frac{d}{d\theta} U[\gamma_{\scriptscriptstyle \mathcal{O}}](\theta; 0) & = - i \left.\left(\frac{dx_{\scriptscriptstyle \mathcal{O}}}{d\theta} K_x + \frac{dy_{\scriptscriptstyle \mathcal{O}}}{d\theta} K_y\right)\right|_{\gamma_{\scriptscriptstyle \mathcal{O}}} U[\gamma_{\scriptscriptstyle \mathcal{O}}](\theta; 0)\\
				& = i r \left.\left(\sin \theta ~ K_x - \cos \theta ~ K_y\right)\right|_{\gamma_{\scriptscriptstyle \mathcal{O}}} U[\gamma_{\scriptscriptstyle \mathcal{O}}](\theta; 0)\\
				& = \frac{- i r \left(r - i \sin \theta\right)}{2 \left(1 - r^2 + 2 i r \sin \theta\right)} \begin{pmatrix}
					0 & - i\\
					i & 0
				\end{pmatrix} U[\gamma_{\scriptscriptstyle \mathcal{O}}](\theta; 0)\\
				& = \frac{i r \left(r - i \sin \theta\right)}{2 \left(1 - r^2 + 2 i r \sin \theta\right)} S \sigma_z S^{-1} U[\gamma_{\scriptscriptstyle \mathcal{O}}](\theta; 0),
			\end{align}
			where $\sigma_z$ is the Pauli Z matrix (for explicit definitions, see Appendix~\ref{Appendix:EvolutionOperators}) and $\displaystyle {S = \frac{1}{\sqrt{2}} \begin{pmatrix} 1 & - i\\ - i & 1 \end{pmatrix}}$.

			By defining 
			\begin{align}
				\Xi_{\scriptscriptstyle \mathcal{O}} (\theta) = S^{-1} U[\gamma_{\scriptscriptstyle \mathcal{O}}](\theta; 0) S,
			\end{align}
			the equation above becomes
			\begin{align}
				\frac{d}{d\theta} \Xi_{\scriptscriptstyle \mathcal{O}} (\theta) = \frac{i r \left(r - i \sin \theta\right)}{2 \left(1 - r^2 + 2 i r \sin \theta\right)} \sigma_z \Xi_{\scriptscriptstyle \mathcal{O}} (\theta).
			\end{align}
			Together with the condition
			\begin{align}
				U[\gamma_{\scriptscriptstyle \mathcal{O}}](0; 0) = \mathds{1} \quad \Leftrightarrow \quad \Xi_{\scriptscriptstyle \mathcal{O}} (0) = \mathds{1},
			\end{align}
			we obtain the solution
			\begin{align}
				\Xi_{\scriptscriptstyle \mathcal{O}} (\theta) = \begin{pmatrix}
					\lambda_{\scriptscriptstyle \mathcal{O}}(\theta) & 0\\
					0 & \lambda_{\scriptscriptstyle \mathcal{O}}(\theta)^{-1}
				\end{pmatrix},
			\end{align}
			where $\lambda_{\scriptscriptstyle \mathcal{O}}(\theta) = \left[\dfrac{\left(1 - r\right) \left(1 + r e^{- i \theta}\right)}{\left(1 + r\right) \left(1 - r e^{- i \theta}\right)}\right]^{\frac{1}{4}}$. Consequently,
			\begin{align}
				U[\gamma_{\scriptscriptstyle \mathcal{O}}](\theta; 0) = \frac{1}{2 \lambda_{\scriptscriptstyle \mathcal{O}}(\theta)} \begin{pmatrix}
					1 + \lambda_{\scriptscriptstyle \mathcal{O}}(\theta)^2 & - i \left(1 - \lambda_{\scriptscriptstyle \mathcal{O}}(\theta)^2\right)\\
					i \left(1 - \lambda_{\scriptscriptstyle \mathcal{O}}(\theta)^2\right) & 1 + \lambda_{\scriptscriptstyle \mathcal{O}}(\theta)^2
				\end{pmatrix}. \label{XiToU}
			\end{align}

			It is worth noting that the real part of the radicand in $\lambda_{\scriptscriptstyle \mathcal{O}}(\theta)$ is always positive for $0 < r <1$. More precisely, the real part of the radicand is 
			\begin{align}
				\Re \frac{\left(1 - r\right) \left(1 + r e^{- i \theta}\right)}{\left(1 + r\right) \left(1 - r e^{- i \theta}\right)} = \Re \frac{\left(1 - r\right) \left(1 - r^2 - 2 i r \sin \theta \right)}{\left(1 + r\right) \left(1 + r^2 - 2 r \cos \theta\right)} = \frac{\left(1 - r\right) \left(1 - r^2 \right)}{\left(1 + r\right) \left(1 + r^2 - 2 r \cos \theta\right)} > 0.
			\end{align}
			Hence, $\lambda_{\scriptscriptstyle \mathcal{O}}(\theta)$ is single-valued, despite the presence of the fourth root. In particular, we can safely conclude that $\lambda(2 \pi) = 1$. Consequently, the evolution operator after completing a loop is
			\begin{align}
				U[\gamma_{\scriptscriptstyle \mathcal{O}}](2 \pi; 0) = \mathds{1}.
			\end{align}
			As a result, the holonomy is trivial; in other words, every state evolves back to itself after being transported along the closed path $\gamma_{\scriptscriptstyle \mathcal{O}}$. This is expected from the fact that the Hilbert space bundle is locally flat.

		\subsection{Holonomy with an EP Enclosed \label{Sec:EnclosingEP}}

			\begin{figure}[h!]
				\begin{center}
					\begin{tabular}{cccc}
						(a) & & (b) &\vspace{-2.5ex}\\
						& \includegraphics[width = 0.35\textwidth]{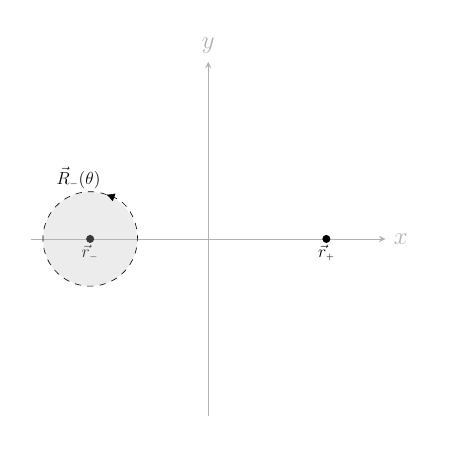} & & \includegraphics[width = 0.35\textwidth]{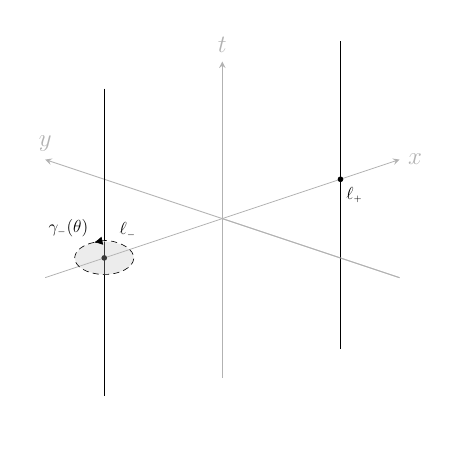}
					\end{tabular}
				\end{center}
				\caption{Transport the state along the path $\gamma_-$. (a) State evolution in the parameter space $\mathds{R}^2$ along ${\vec{R}_- (\theta) = \left(-1 + \rho \cos \theta\right) ~ \hat{e}_x + \rho \sin \theta ~ \hat{e}_y}$, which is the projection of $\gamma_-$ onto the parameter space. The EP at $\vec{r}_-$ is enclosed by the loop. (b) State evolution in the base space $\mathds{R} \times \mathds{R}^2$. The closed path $\gamma_-$ encircles the line $\ell_-$ at $t = 0$, which is traced out in the base space by the EP $\vec{r}_-$ in the parameter space.}
				\label{Fig:EncircleEP}
			\end{figure}

			Despite evolving during transport along $\gamma_{\scriptscriptstyle \mathcal{O}}$, the state returns to its initial form at the end of the loop. However, this does not necessarily mean that a state will evolve back to itself after completing every closed-path evolution. In fact, if the path encloses an EP, the state generally does not return to its original form.

			To be more specific, as mentioned in Sec.~\ref{Sec:Introduction}, numerous studies have demonstrated that encircling an EP causes eigenvalues to traverse between connected Riemann sheets~\cite{Szekeres1959, Dembowski2004, Zhong2018, Oezdemir2019, Arkhipov2023, Arkhipov2024, Lai2024, Beniwal2024, Roy2025, Kawabata2019, Hu2022, Yang2022, Okuma2023, Jia2023}. Because eigenvalues are directly associated with eigenstates, which are themselves special cases of general quantum states, this transition strongly indicates that the holonomies surrounding an EP can be nontrivial.

			To show this explicitly, we find the evolution operator along the path $\gamma_-: [0, 2 \pi] \to \mathds{R} \times \mathds{R}^2$ (see Fig.~\ref{Fig:EncircleEP}), where $\gamma_-$ is parameterized as
			\begin{align}
				\gamma_- (\theta) = (0, x_- (\theta), y_- (\theta)) = (0, - 1 + \rho \cos \theta, \rho \sin \theta),
			\end{align}
			where $0 < \rho < 2$, so that only the EP $\ell_-$ is enclosed by the closed path $\gamma_-$.

			The evolution operator $U[\gamma_-](0; 0)$ along this path $\gamma_-$ is then governed by
			\begin{align}
				\frac{d}{d\theta} U[\gamma_-](\theta; 0) & = - i \left.\left(\frac{dx_-}{d\theta} K_x + \frac{dy_-}{d\theta} K_y\right)\right|_{\gamma_-} U[\gamma_-](\theta; 0)\\
				& = i \rho \left.\left(\sin \theta ~ K_x - \cos \theta ~ K_y\right)\right|_{\gamma_-} U[\gamma_-](\theta; 0)\\
				& = \frac{- \left(1 - \rho e^{- i \theta}\right)}{2 \left(2 - \rho e^{- i \theta}\right)} \sigma_y U[\gamma_-](\theta; 0)\\
				& = \frac{\left(1 - \rho e^{- i \theta}\right)}{2 \left(2 - \rho e^{- i \theta}\right)} S \sigma_z S^{-1} U[\gamma_-](\theta; 0). \label{UGammaMinus}
			\end{align}

			Once again, we let
			\begin{align}
				\Xi_- (\theta) = S^{-1} U[\gamma_-](\theta; 0) S,
			\end{align}
			together with the condition $U[\gamma_-](0; 0) = \mathds{1}$ (i.e., $\Xi_- (0) = \mathds{1}$), we obtain
			\begin{align}
				\Xi_- (\theta) = \begin{pmatrix}
					\lambda_-(\theta) & 0\\
					0 & \lambda_-(\theta)^{-1}
				\end{pmatrix},
			\end{align}
			where $\lambda_-(\theta) = e^{- i \theta / 4} \left(\dfrac{2 - \rho e^{- i \theta}}{2 - \rho}\right)^{1 / 4}$. Clearly, as shown previously in Eq.~\eqref{XiToU},the evolution operator is
			\begin{align}
				U[\gamma_-](\theta; 0) = \frac{1}{2 \lambda_-(\theta)} \begin{pmatrix}
					1 + \lambda_-(\theta)^2 & - i \left(1 - \lambda_-(\theta)^2\right)\\
					i \left(1 - \lambda_-(\theta)^2\right) & 1 + \lambda_-(\theta)^2
				\end{pmatrix}.
			\end{align}

			Interestingly, since
			\begin{align}
				\Re \frac{2 - \rho e^{- i \theta}}{2 - \rho} = \frac{2 - \rho \cos \theta}{2 - \rho} > 0,
			\end{align}
			for $0 < \rho < 2$, $\lambda_-(\theta)$ is also single-valued. Nevertheless, $\lambda_-(2 \pi) = - i \neq 1$. Hence,
			\begin{align}
				U[\gamma_-](2 \pi; 0) = \begin{pmatrix}
					0 & 1\\
					- 1 & 0
				\end{pmatrix} =: \mathcal{I}, \label{UEP}
			\end{align}
			so that states do not evolve back to themselves after completing the path $\gamma_-$ in general. Rather, they need to wind around the EP four times in order to return to their original configuration.

			To show that this effect indeed arises from encircling the line $\ell_-$ swept out by the EP, we note that $U[\gamma_-](2 \pi; 0)$ in Eq.~\eqref{UEP} does not depend on the radius $\rho$ of the path $\gamma_-$ enclosing the line $\ell_-$. Therefore, the path can be taken arbitrarily close to $\ell_-$ (i.e., $\rho \rightarrow 0^+$) without changing the fact that the states fail to recover their initial configurations upon completing the loop. Therefore, $\ell_-$ becomes a topological defect in the base space.

			\begin{figure}[h!]
				\begin{center}
					\begin{tabular}{cccc}
						(a) & & (b) &\vspace{-2.5ex}\\
						& \includegraphics[width = 0.35\textwidth]{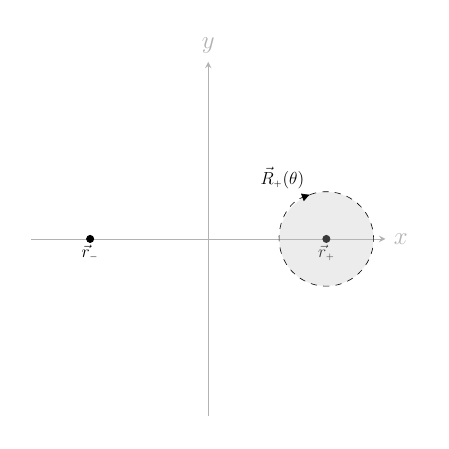} & & \includegraphics[width = 0.35\textwidth]{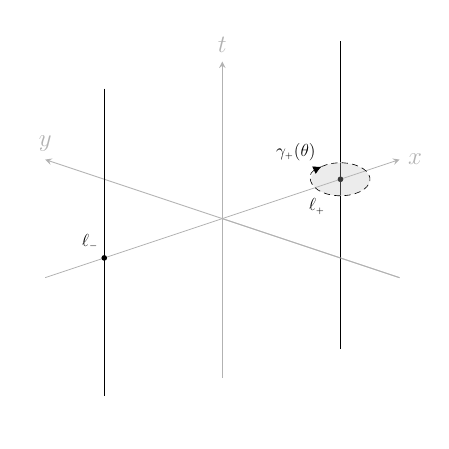}
					\end{tabular}
				\end{center}
				\caption{Transport the state along the path $\gamma_+$. (a) State evolution in the parameter space $\mathds{R}^2$ along ${\vec{R}_+ (\theta) = \left(1 - \rho \cos \theta\right) ~ \hat{e}_x + \rho \sin \theta ~ \hat{e}_y}$, i.e., a path encircles the EP at $\vec{r}_+$. (b) State evolution in the base space $\mathds{R} \times \mathds{R}^2$. The closed path $\gamma_+$ encircles the line $\ell_+$ at $t = 0$.}
				\label{Fig:EncircleAnotherEP}
			\end{figure}

			Regarding the other line $\ell_+$ swept out by the EP at $\vec{r}_+$, we can enclose it by the path $\gamma_+: [0, 2 \pi] \to \mathds{R} \times \mathds{R}^2$ (see Fig.~\ref{Fig:EncircleAnotherEP}) parameterized by
			\begin{align}
				\gamma_+ = (0, x_+, y_+) = (0, 1 - \rho \cos \theta, \rho \sin \theta).
			\end{align}
			This path, in fact, leads to the following evolution operator equation:
			\begin{align}
				\frac{d}{d\theta} U[\gamma_+](\theta; 0) = - \frac{\left(1 - \rho e^{- i \theta}\right)}{2 \left(2 - \rho e^{- i \theta}\right)} S \sigma_z S^{-1} U[\gamma_+](\theta; 0),
			\end{align}
			which is almost the same equation as Eq.~\eqref{UGammaMinus}, except for an overall minus sign. Hence, we can deduce that
			\begin{align}
				U[\gamma_+](2 \pi; 0) = \mathcal{I}^{-1} = \begin{pmatrix}
					0 & - 1\\
					1 & 0
				\end{pmatrix},
			\end{align}
			without repeating all the steps for path $\gamma_+$. Alternatively, we can simply use the fact that $x_+ = - x_-$ and $H(-x, y) = T H(x, y) T$, where $T = \begin{pmatrix} 0 & 1\\1 & 0 \end{pmatrix}$, so that $U[\gamma_+](2 \pi; 0) = T U[\gamma_-](2 \pi; 0) T = \mathcal{I}^{-1}$ is expected. In other words, winding around $\gamma_+$ or $\gamma_-$ once changes the state $\ket{\Psi}$ by a factor of $\mathcal{I}^{\pm 1}$, depending on the orientation.

		\subsection{Holonomy with Nontrivial Time Loop \label{Sec:TimeEvolution}}

			In the previous cases, we focused on the holonomies on a specific time slice ($t = 0$). To further demonstrate that the nontrivial holonomies are caused by the line swept out by the EPs, we demonstrate a holonomy using curves with explicit time dependence.

			In this specific example, we consider a path $\widetilde{\gamma}_-:[0, 2\pi] \rightarrow \mathds{R} \times \mathds{R}^2$ parameterized by
			\begin{align}
				\widetilde{\gamma}_- (\theta) & = (T (\theta), x_- (\theta), y_- (\theta)) = \left(T(\theta), -1 + \rho \cos \theta, \rho \sin \theta\right),
			\end{align}
			where $T(\theta)$ satisfies $T(0) = T(2\pi) = 0$. Although $T(\theta)$ could be piecewise smooth, we restrict our attention to a smooth function for simplicity, which ensures that $\widetilde{\gamma}_-$ forms a closed loop.

\begin{figure}[h!]
				\begin{center}
					\includegraphics[width = 0.35\textwidth]{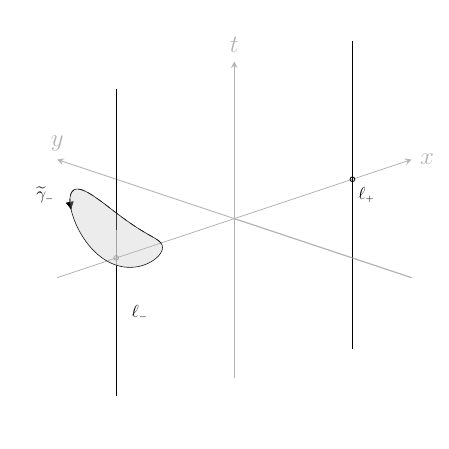}
				\end{center}
				\caption{Transport of the state along a time-dependent closed loop $\widetilde{\gamma}_-$ enclosing $\ell_-$ in the base space $\mathds{R} \times \mathds{R}^2$. The resulting holonomy (closed-loop evolution) is independent of time.}
				\label{Fig:NontrivialTimeLoop}
			\end{figure}

			It is worth noting that despite the fact that function $T (\theta)$ is arbitrary, $\widetilde{\gamma}_-$ is a loop enclosing line $\ell_{\scriptscriptstyle -}$ (see Fig.~\ref{Fig:NontrivialTimeLoop} for an illustration of this loop).

			Following Eq.~\eqref{EvolutionEq}, the evolution operator $U[\widetilde{\gamma}_-]$ is governed by
			\begin{align}
				\frac{d}{d \theta} U[\widetilde{\gamma}_-](\theta; 0) & = - i \left.\left(\frac{d T}{d\theta} H + \frac{dx_-}{d\theta} K_x + \frac{dy_-}{d\theta} K_y \right)\right|_{\gamma_1} U[\widetilde{\gamma}_-](\theta; 0)\\
				& = - i \left.\left(T' H - \rho \sin \theta K_x + \rho \cos \theta K_y \right)\right|_{\gamma_1} U[\widetilde{\gamma}_-](\theta; 0), \label{Utildegamma}
			\end{align}
			with $U[\widetilde{\gamma}_-](0; 0) = \mathds{1}$, where $T' \equiv \dfrac{d T}{d \theta}$.

			To prevent the detailed derivation from distracting from the main logic flow, we defer the solution of Eq.~\eqref{Utildegamma} to Appendix~\ref{Appendix:EvolutionOperators}, presenting only the final result here:
			\begin{align}
				U[\widetilde{\gamma}_-](\theta; 0) = e^{- i T(\theta) H(x_-(\theta), y_-(\theta))} U[\gamma_-](\theta; 0),
			\end{align}
			where is $H(x_-(\theta), y_-(\theta))$ the Hamiltonian evaluated at $x_-(\theta)$ and $y_-(\theta)$, and $U[\gamma_-](\theta; 0)$ is the evolution operator on the $t = 0$ slice defined in Sec.~\ref{Sec:EnclosingEP}.

\begin{figure}[h!]
				\begin{center}
					\includegraphics[width = 0.35\textwidth]{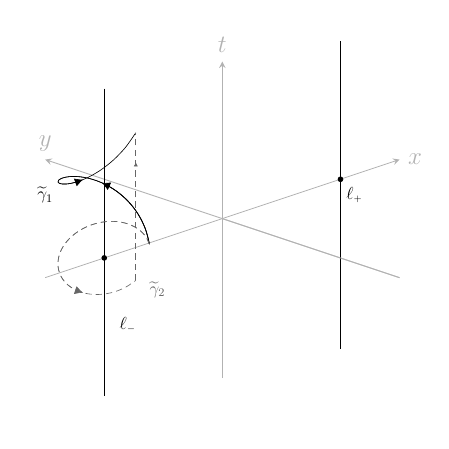}
				\end{center}
				\caption{The evolution operator along path $\widetilde{\gamma}_1$ is identical to that along path $\widetilde{\gamma}_2$, where both paths lie in the base space $\mathds{R} \times \mathds{R}^2$, where the latter proceeds along the $t = 0$ slice before evolving solely in time (see Appendix~\ref{Appendix:EvolutionOperators} for the explicit calculation). This result is expected, as no obstructions exist between the two paths.}
				\label{Fig:Equivalence}
			\end{figure}

			This result is, in fact, easy to picture. The $U[\gamma_-](\theta; 0)$ term says that it evolves from $0$ to $\theta$ on the $t = 0$ plane, after which the system undergoes additional evolution governed by the Hamiltonian evaluated at $(x_-(\theta), y_-(\theta))$ for a duration of $T(\theta)$ (see Fig.~\ref{Fig:Equivalence} for an illustration).

			Therefore, the holonomy along the nontrivial time loop $\widetilde{\gamma}_-$ encircling $\ell_-$ is
			\begin{align}
				U[\widetilde{\gamma}_-](2 \pi; 0) & = e^{- i T(2 \pi) H(x_-(\theta), y_-(\theta))} U[\gamma_-](2 \pi; 0)\\
				& = U[\gamma_-](2 \pi; 0) = \mathcal{I},
			\end{align}
			which is precisely the holonomy found on the $t = 0$ slice.

			This ``path-independent'' result is, in fact, consistent with the ``locally flat'' property of the Hilbert space bundle, namely, that only the endpoints of the path matter, provided no topological defects are encountered.

			To summarize, the local flatness of the Hilbert space bundle, together with the presence of nontrivial holonomy around $\ell_-$ and $\ell_+$, suggests that $\ell_-$ and $\ell_+$ serve as localized sources of topological phase. The emergence of nontrivial holonomy upon encircling the line swept out by an EP suggests that the EP acts as a topological defect. These facts suggest that the parameter space is $\mathcal{M}^2 = \mathds{R}^2 \setminus \left\{\vec{r}_\pm\right\}$, which implies that the base space is $\mathds{R} \times \mathds{R}^2 \setminus \left(\ell_- \cup \ell_+\right)$ rather than the naive guess of $\mathds{R} \times \mathds{R}^2$. Consequently, this topological structure implies that the holonomy group contains at least a $\mathds{Z}_4$ subgroup, as indicated by $\mathcal{I}^4 = \mathds{1}$.

	\section{Physical Signatures of the Topological Defect}

		Based on the discussion in the previous sections, EPs can be detected using the direct calculation of holonomies, which involves closed loops in $(\mathds{R} \times \mathcal{M}^n)\setminus E$. Here, $\mathds{R}$ corresponds to the timeline, $\mathcal{M}^n$ is the $n$-dimensional parameter space, and $E$ is the defect caused by the EP (in our example above, $\mathcal{M}^n = \mathds{R}^2$ and $E = \ell_+ \cup \ell_-$, where $\ell_\pm = \{(t, \vec{r}) \in \mathds{R} \times \mathds{R}^2 \mid \vec{r} = \vec{r}_\pm\}$).

		Nevertheless, this ``detection'' is largely abstract, theoretical, or purely mathematical. In a real physical system, parameter modulation requires finite time, making the physical realization of a closed loop in the time-parameter space unrealistic, if not impossible. Consequently, it might appear that the concept of holonomy holds no practical utility and that the underlying topological defect is of little real-world consequence.

		This can be done because the emergence of parameter evolution generators $K$ is a direct consequence of the mathematical properties of the metric $G$ and the assumption that physical states remain physical [most importantly, normalizability in Eq.~\eqref{Assumption}]~\cite{Ju2024}. In other words, no actual physical processes or approximations are involved in the formulation. To make the discussion more compatible with realistic physical systems, we should also take time into consideration, like what has been done in Sec.~\ref{Sec:TimeEvolution}, by introducing a linear dependence on time.

		Specifically, if a system depends on $n$ parameters besides time, say, $\{q^1, q^2, \dots, q^n\}$, there should be corresponding $n$ parameter evolution generators $\{K_1, K_2, \dots, K_n\}$. Together with time $T$ and the time-evolution generator $H$, the governing equation [Eq.~\eqref{EvolutionEq}] for the evolution operator along a specific path $\gamma$ can be written as
		\begin{align}
			i \frac{d}{d s} U[\gamma] (s; s_0) = \left(\frac{d T}{d s} H + \frac{d q_\gamma^i}{d s} K_i\right) U[\gamma] (s; s_0),
		\end{align}
		where $\{T, q_\gamma^i\}$ are the parameterization components of the path and are functions of the path parameter $s$, and $s_0$ is the initial point of the path parameter. By choosing the path parameter to be time itself ($s = t$), which identifies the time component as $T = t$, the equation simplifies to the evolution equation
		\begin{align}
			i \frac{d}{d t} U[\gamma] (t; t_0) = \left(H + v^i K_i\right) U[\gamma] (t; t_0), \label{PhysicalEvolutionEq}
		\end{align}
		where $v^i = \dfrac{d q_\gamma^i}{d t}$ is the rate of change of the parameters (or the ``velocity'' in the parameter space $\mathcal{M}^n$). The expression in Eq.~\eqref{PhysicalEvolutionEq} provides a more transparent representation of the rate at which the parameters change.

		\begin{figure}[h!]
			\begin{center}
				\includegraphics[width = 0.35\textwidth]{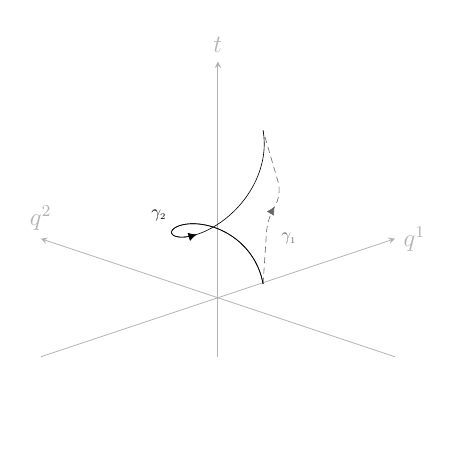}
			\end{center}
			\caption{Although the paths $\gamma_1$ and $\gamma_2$ are very different in the base space $\mathds{R} \times \mathds{R}^2$, the evolution along the two paths is identical because their endpoints are the same, no EP (topological defect) lies between them, and the Hilbert space bundle is locally flat.}
			\label{Fig:Different}
		\end{figure}

		\begin{figure}[h!]
			\begin{center}
				\includegraphics[width = 0.35\textwidth]{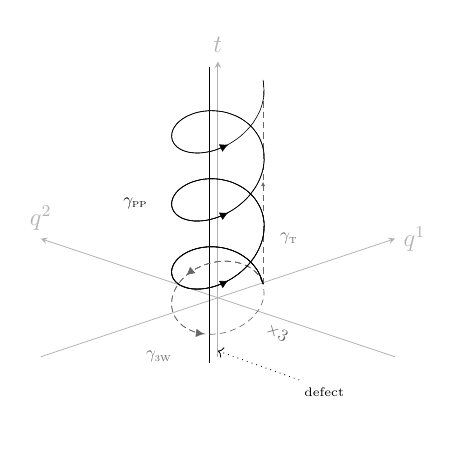}
			\end{center}
			\caption{In a real physical system, parameter changes require finite time. Consequently, a physical path evolves in the time direction and generally does not form a closed loop. However, as illustrated, a physical path $\gamma_{\text{\tiny PP}}$ winding three times around the EP is equivalent to a 3-winding holonomy ($\gamma_{\text{\tiny 3W}}$) followed by a pure time evolution $\gamma_\text{T}$ (where all paths lie in the base space $\mathds{R} \times \mathds{R}^2$). This total evolution differs fundamentally from a trivial, pure time evolution due to the enclosure of the EP, highlighting the crucial role that holonomy plays in the full dynamics.}
			\label{Fig:Winding}
		\end{figure}

		However, as demonstrated in Sec.~\ref{Sec:TimeEvolution}, because the Hilbert space bundle is locally flat everywhere, any two paths sharing the same initial and final points with no topological defects lying between them can be continuously deformed into one another, resulting in identical time evolutions (see Fig.~\ref{Fig:Different} for illustration). In contrast, a quantum state encircling an EP (a topological defect) cannot be continuously unwound. Consequently, the resulting time evolution generally depends on the winding number and its holonomy even though the path does not form a closed loop in a mathematical sense (see Fig.~\ref{Fig:Winding} for illustration). More precisely, the final state resulting from winding around an EP is generally different from a trivial evolution that does not encircle the defect.

		In other words, we can still ``detect'' the effect from the nontrivial holonomy caused by the EPs in experiments. To be more specific, to determine if an EP lies between two evolutionary paths, simply evolve a quantum state along both paths from the same starting point to the same final endpoint; if subsequent measurements reveal different resulting states, a topological defect is guaranteed to be enclosed between them.

		On the other hand, by further understanding holonomies induced by the EPs, we can potentially utilize their topological properties to precisely manipulate and control quantum states.

	\section{Conclusion}

		Rather than limiting ourselves to an eigenstate, which form a subspace of the whole space, we examine the topology of the entire Hilbert space bundle. Specifically, we demonstrate that an arbitrary quantum state traveling along a closed path does not necessarily return to the same state if the path encloses an EP; hence, the Hilbert space bundle with EPs admits nontrivial holonomies. Given that the bundle is locally flat, the presence of nontrivial holonomies suggests the existence of underlying topological defect.

		Furthermore, by considering the full Hilbert space bundle, our framework incorporates all possible state configurations by construction. This eliminates the need for special treatments, such as non-Abelian Berry connections or domain enlargement, and provides a unified description of the evolution of quantum states, including around EPs, without requiring prior identification of their presence or order.

		Upon examining the holonomies more closely, we not only find that EPs can lead to nontrivial holonomy, but also identify that this nontrivial holonomy specifically arises from encircling an EP. Although a 2nd-order EP is explicitly demonstrated in this work, the movement of eigenvalues across Riemann sheets when encircling EPs provides strong evidence that this behavior also occurs for higher-order EPs, as discussed in Sec.~\ref{Sec:EnclosingEP}. Therefore, EPs effectively act as topological defects, inducing exotic behavior in quantum states when they are transported around these points.

		This study demonstrates that EPs can serve as topological defects, offering an explicit example of a 2nd-order EP where four complete windings are required for a quantum state to return to its initial form. However, while the required number of windings for general Hamiltonian eigenstates has been discussed in literature~\cite{Guria2024, Ju2025b}, this study cannot determine how many windings are required for a general quantum state around an arbitrary EP. Consequently, a comprehensive classification of how EPs govern state topology remains to be established.

		This phenomenon is not only of theoretical interest, but could also provide some practical applications. One potential application is leveraging nontrivial holonomies to implement quantum gates and manipulate qubits.

		Therefore, this study marks the beginning of a broader exploration. Many important questions remain open and await further investigation.

	\section*{Acknowledgments}
	
		C.Y.J. is partially supported by the National Science and Technology Council through Grant No. NSTC 112-2112-M-110-013-MY3, NSTC 114-2112-M-029-003, and NSTC 115-2112-M-110-011.
		
	\section*{Author Contributions}
	
		C.Y.J. conceptualized and initialized the idea. Both C.Y.J. and S.M.C performed the detailed calculations and contributed to the preparation and writing of the manuscript. Additionally, a Large Language Model (Gemini) was used to provide language support, and the resulting text was subsequently reviewed, edited, and approved by all authors, who take full responsibility for the integrity and content of the work.
	
	\appendix
	\section*{Appendices}
	\setcounter{equation}{0}
	\renewcommand{\theequation}{\thesection.\arabic{equation}}

		\section{Berry Connections from Projections}

			The Berry connection is a standard tool for understanding how states depend on parameters. Nevertheless, a Berry connection carries information about the parameter evolution of a specific Hamiltonian eigenstate, which can be viewed as a subspace of the full Hilbert space. In this Appendix, we show that the Berry connection can, in general, be obtained by projecting the connection of the full Hilbert space bundle onto the eigenstate under a specific gauge choice appearing in Eqs.~\eqref{AdiabaticGauge} and \eqref{AdiabaticGauges} (discussed in detail in Appendix~\ref{Appendix:AdiabaticGauge}).

			Under the adiabatic approximation, we consider a quantum system whose full quantum state $\ket{\Psi_n (t, \{q\})}$ evolves slowly as a set of Hamiltonian parameters $\left\{q^1, q^2, \dots, q^n\right\}$ (where $q^i$ denotes the $i$th parameter, not the $i$th power of $q$) varies with time. Assuming the parameter rates of change are negligible $\left(\dfrac{d q^i}{d t} \approx 0\right)$, the state factorizes into a dynamical phase, a Berry phase, and a Hamiltonian eigenstate,
		\begin{align}
			\ket{\Psi_n (t, \{q\})} = e^{i \gamma_n [\mathcal{C}]}e^{- i \theta_n(t)} \ket{\psi_n(\{q\})},
		\end{align}
		where $\displaystyle \theta_n (t) = \int_{t_0}^t h_n(\{q(t')\}) dt'$ is the dynamical phase factor, $\gamma_n [\mathcal{C}]$ is the Berry phase along path $\mathcal{C}$, $\ket{\psi_n(\{q\})}$ is an instantaneous eigenstate of the Hamiltonian $H(\{q\})$ satisfying the time-independent eigenvalue equation
		\begin{align}
			H (\{q\}) \ket{\psi_n(\{q\})} = h_n (\{q\}) \ket{\psi_n(\{q\})}
		\end{align}
		with $h_n (\{q\})$ being the corresponding energy eigenvalue. In the Hermitian regime, the Berry connection associated with state $n$ is given by
		\begin{align}
			\p{\mathcal{A}}{n}_i (\{q\}) = i \bra{\psi_n(\{q\})} \partial_i \ket{\psi_n(\{q\})},
		\end{align}
		where $\partial_i = \partial / \partial q^i$.

			With this definition, the Berry phase mentioned above can be written simply as
		\begin{align}
			\gamma_n [\mathcal{C}] = \int_\mathcal{C} dq^i \p{\mathcal{A}}{n}_i,
		\end{align}
		where $\mathcal{C}$ is a path in the parameter space.

			For conciseness, we will omit the explicit parameter dependence $(t, \{q\})$ where no confusion arises.

			Nevertheless, when it comes to the non-Hermitian regime, ambiguity arises. In classical or semiclassical systems, non-Hermiticity implies amplification or attenuation, which leads to a complex Berry phase where the imaginary part is related to the gain or decay of the state~\cite{Silberstein2020, Singhal2023, Ozawa2025, Montag2026, Lane2025}. However, in the fully quantum regime, the norms of quantum states should be preserved.

			Therfore, to generalize the Berry connection to the non-Hermitian quantum regime, both left- and right-eigenstates~\cite{Brody2013} are required in its construction~\cite{Garrison1988, MehriDehnavi2008, Shen2018}, namely,
		\begin{align}
			\p{\mathcal{A}}{n}_i = i ~ \lbra{\psi_n} \partial_i \rket{\psi_n}, \label{App:BerryConnection}
		\end{align}
		where
		\begin{align}
			\lbra{\psi_n} H = h_n \lbra{\psi_n}, \quad \text{and} \quad H \rket{\psi_n} = h_n \rket{\psi_n}.
		\end{align}
		This leads to the cancellation of the phases, which renders the norm invariant~\cite{Arkhipov2026}.

			In fact, as will be shown later, the Berry connection can also be obtained from
		\begin{align}
			\p{\mathcal{A}}{n}_i = \Bra{\psi_n} K_i \Ket{\psi_n}, \label{App:BerryConnectionAsProjection}
		\end{align}
		where $K_i$ is the parameter generator in the $q^i$-direction under the gauge condition given in Eq.~\eqref{AdiabaticGauges}.

			The detailed derivation of this result, along with its geometric implications, is presented across the following subsections.

			\subsection{Adiabatic Gague \label{Appendix:AdiabaticGauge}}

				To derive the Berry connection in the form of Eq.~\eqref{App:BerryConnectionAsProjection}, we first review the gauge condition given in Eq.~\eqref{AdiabaticGauges},
			\begin{align}
				[\partial_t K_i, H] = 0,
			\end{align}
			following Ref.~\cite{Ju2024}. (See Appendix~\ref{Appendix:GaugeDOF} for a review of gauge symmetry in the Hilbert space bundle.)

				However, because Eq.~\eqref{AdiabaticGauges} represents a multi-parameter generalization of the single-parameter condition in Eq.~\eqref{AdiabaticGauge},
			\begin{align}
				[\partial_t K, H] = 0,
			\end{align}
			we focus first on the single-parameter case for simplicity. Its extension to multiple parameters follows straightforwardly within this framework.

				On the technical side, as derived in the main text, not only does this gauge condition render the $q$-evolution generator $K$ linear in time $t$,
			\begin{align}
				K = t \p{K}{1} + \p{K}{0},
			\end{align}
			but it also reduces the governing equation for $K$ from a differential equation into a set of algebraic equations, which reads
			\begin{align}
				0 = \partial_t K - \partial_q H + i \left[H, K\right] \quad \Rightarrow \quad \left\{\begin{array}{l}
					\displaystyle \left[\p{K}{1}, H\right] = 0\\
					\displaystyle \p{K}{1} + i \left[H, \p{K}{0}\right] = \partial_q H
				\end{array}\right.. \label{App:DiffToAlge}
			\end{align}
			In other words, this gauge condition is particularly easy to work with.

				It is worth emphasizing that $\p{K}{0}$ in Eq.~\eqref{App:DiffToAlge} is not uniquely determined. Adding a $\Delta K$ to $\p{K}{0}$ yields another valid solution to Eq.~\eqref{App:DiffToAlge}, provided that
			\begin{align}
				[\Delta K, H] = 0. \label{App:ResidualGauge}
			\end{align}
			This result turns out to correspond to the gauge transformation of the Berry connection that will be discussed later.

				Beyond these technical advantages listed in Eq.~\eqref{App:DiffToAlge}, this gauge choice aligns naturally with the Berry connection. As noted at the beginning of this Appendix, the Berry connection is constructed from derivatives of Hamiltonian eigenstates. More precisely, it measures the overlap between the same eigenstate evaluated at two nearby points in parameter space during an adiabatic process.

				Coincidentally, this gauge choice is compatible with adiabatic evolution. To see this, we begin with the equation governing the parameter evolution of a quantum state,
			\begin{align}
				\partial_q \ket{\Psi} = - i K \ket{\Psi}.
			\end{align}
			To ensure that evolution in the $q$-direction is compatible with adiabatic evolution, let $\ket{\Psi}$ be the $n$th eigenstate of the Hamiltonian $H(q)$ with eigenvalue $h_n(q)$, i.e.,
			\begin{align}
				H \ket{\Psi_n} = h_n \ket{\Psi_n}
			\end{align}
			for all possible $(t, q)$.

				Taking the $q$-derivative on both sides yields
			\begin{align}
				& \quad \partial_q \big( H \ket{\Psi_n} \big) = \partial_q \big( h_n \ket{\Psi_n} \big)\\
				\Rightarrow & \quad \left(\partial_q H \right) \ket{\Psi_n} + H \partial_q \left(\ket{\Psi_n}\right) = \left(\partial_q h_n\right) \ket{\Psi_n} + h_n \left(\partial_q \ket{\Psi_n}\right)\\
				\Rightarrow & \quad \left(\partial_q H - i H K \right) \ket{\Psi_n} = \left(\partial_q h_n - i h_n K\right) \ket{\Psi_n}\\
				\Rightarrow & \quad \left(\partial_q H - i H K \right) \ket{\Psi_n} = \left(\partial_q h_n - i K H\right) \ket{\Psi_n}\\
				\Rightarrow & \quad \left(\partial_q H + i \left[ K , H \right] \right) \ket{\Psi_n} = \left(\partial_q h_n\right) \ket{\Psi_n}\\
				\Rightarrow & \quad \left(\partial_t K \right) \ket{\Psi_n} = \left(\partial_q h_n \right) \ket{\Psi_n}.
			\end{align}
			That is, $\ket{\Psi_n}$ is not only an eigenstate of $H$ with eigenvalue $h_n$, but also an eigenstate of $\partial_t K = \p{K}{1}$ with eigenvalue $\partial_q h_n$.

				Furthermore, if adiabatic evolution holds for every $n$, this implies that
			\begin{align}
				\left[\partial_t K , H \right] = 0, \label{App:AdiabaticGauge}
			\end{align}
			which is exactly the gauge condition appeared in Eq.~\eqref{AdiabaticGauge}.

				On the other hand, multiplying an eigenstate by an arbitrary non-zero function yields another eigenstate. In other words, replacing $\ket{\Psi_n}$ with $\ket{\widetilde{\Psi}_n} = f_n \ket{\Psi_n}$ does not alter the condition in Eq.~\eqref{App:AdiabaticGauge}. However, it modifies the explicit expression for $\p{K}{0}$. . To be specific, setting
			\begin{align}
				\partial_q \ket{\Psi_n} = - i K \ket{\Psi_n},
			\end{align}
			gives
			\begin{align}
				& \partial_q \ket{\widetilde{\Psi}_n} = - i \widetilde{K} \ket{\widetilde{\Psi}_n}\\
				\Rightarrow & \left(\partial_q f_n\right) \ket{\Psi_n} + f_n \left(\partial_q \ket{\Psi_n}\right) = - i \widetilde{K}\left(f_n \ket{\widetilde{\Psi}_n}\right)\\
				\Rightarrow & \left(\partial_q f_n\right) \ket{\Psi_n} - i f_n K \ket{\Psi_n} = - i f_n \widetilde{K}\ket{\Psi_n}\\
				\Rightarrow & K \ket{\Psi_n} - i \left(\partial_q \ln f_n\right) \ket{\Psi_n} = \widetilde{K}\ket{\Psi_n}.
			\end{align}
			Considering all states $n$ simultaneously, we find
			\begin{align}
				\widetilde{K} = K + \sum_n \zeta_n \ket{\Psi_n}\bra{\Phi_n}, \label{App:ExplicitResidualGauge}
			\end{align}
			where $\zeta_n = i \ln f_n$ and $\bra{\Phi_n}$ is a left eigenstate of $H$ satisfying $\braket{\Phi_m}{\Psi_n} = \delta_{mn}$. In other words, for any arbitrarily chosen $C_n$, the second term on the right-hand side of Eq.~\eqref{App:ExplicitResidualGauge} satisfies
			\begin{align}
				\left[\sum_n \zeta_n \ket{\Psi_n}\bra{\Phi_n}, H\right] & = \sum_n \zeta_n \ket{\Psi_n}\bra{\Phi_n} H - \sum_n \zeta_n H \ket{\Psi_n}\bra{\Phi_n}\\
				& = \sum_n \zeta_n (h_n - h_n) \ket{\Psi_n}\bra{\Phi_n} = 0.
			\end{align}
			This result corresponds directly to $\Delta K$ appearing in Eq.~\eqref{App:ResidualGauge}.

				For a system parameterized by $n$-dimensional parameters $\{q^1, q^2, \dots, q^n\}$, the preceding 1D derivation generalizes naturally to each coordinate component $q^i$. By applying the parameter derivative $\partial_i$ to the eigenvalue equation $H\ket{\Psi_n} = h_n\ket{\Psi_n}$ and defining the corresponding generator $\partial_i \ket{\Psi_n} = - i K_i\ket{\Psi_n}$, an identical sequence of commutator manipulations recovers Eq.~\eqref{AdiabaticGauges}, namely,
			\begin{align}
				[\partial_t K_i, H] = 0 \quad \text{for all } i \in {1, \dots, n}.
			\end{align}
			Thus, each component $K_i$ independently satisfies the gauge requirement along its respective parameter direction.

				Hence, the gauges provided in Eqs.~\eqref{AdiabaticGauge} and \eqref{AdiabaticGauges} are compatible with the discussion of the Berry connections.

			\subsection{Berry Connection from Evolution Generators}

				Having established that the evolution generator $K_i$ under the gauge condition in Eq.~\eqref{AdiabaticGauges} is compatible with the adiabatic process, the relation between the Berry connection and the evolution generator in the full Hilbert space bundle arises naturally.

				By defining $G = \sum_n \ket{\Phi_n}\bra{\Phi_n}$, it is straightforward to verify that it satisfies Eq.~\eqref{DeriveMetricQEq}, namely,
			\begin{align}
				\partial_t G & = \partial_t \sum_n \ket{\Phi_n}\bra{\Phi_n} = \sum_n \left[\left(\partial_t \ket{\Phi_n}\right)\bra{\Phi_n} + \ket{\Phi_n}\left(\partial_t \bra{\Phi_n}\right)\right]\\
				& = \sum_n - i \left[\left(H^\dagger \ket{\Phi_n}\right)\bra{\Phi_n} + i \ket{\Phi_n}\left(\bra{\Phi_n} H\right)\right]\\
				& = - i H^\dagger \left(\partial_t \sum_n \ket{\Phi_n}\bra{\Phi_n}\right) + i \left(\partial_t \sum_n \ket{\Phi_n}\bra{\Phi_n}\right) H\\
				& = - i H^\dagger G + i G H,
			\end{align}
			where $\bra{\Phi_n}$ is the $n$th eigenstate satisfying $\braket{\Phi_m}{\Psi_n} = \delta_{mn}$ with $\ket{\Phi_n} = \bra{\Phi_n}^\dagger$ as mentioned in the previous section. Consequently, $G$ serves as a metric for the Hilbert space bundle.

				On the other hand, since $\ket{\Psi_n}$ is a right eigenstate of the Hamiltonian, the state $\rket{n}$ in Eq.~\eqref{App:BerryConnection} can be chosen as $\rket{n} = \ket{\Psi_n}$, which renders $\Bra{\Psi_n} = \bra{\Psi_n} G = \bra{\Phi_n}$. Consequently, Eq.~\eqref{App:BerryConnection} can be written as
			\begin{align}
				\p{\mathcal{A}}{n}_i & = i ~ \lbra{\psi_n} \partial_i \rket{\psi_n}\\
				& = i ~ \Bra{\psi_n} \partial_i \Ket{\psi_n}\\\
				& = \Bra{\psi_n} K_i \Ket{\psi_n}.
			\end{align}

				A subtle point: $K_i$ in the last expression is evaluated at $t = t_0$, which is $K_i = t_0 \p{K}{1} + \p{K}{0}$. However, because $[\p{K}{1}, H] = 0$, $t_0 \p{K}{1}$ acts as a pure gauge contribution and can be eliminated via a gauge transformation, namely $\Delta K = - t_0 \p{K}{1}$. This will become clear in the next subsection during the discussion of the Berry curvature.

				Alternatively, since
			\begin{align}
				\mathds{P}_n = \Ket{\psi_n} \Bra{\psi_n}
			\end{align}
			is the projection operator onto the $n$th eigenstate, the projection of the operator $K_i$ onto the $n$th eigenstate is given by
			\begin{align}
				\mathds{P}_n K_i \mathds{P}_n = \Ket{\psi_n} \Bra{\psi_n} K_i \Ket{\psi_n} \Bra{\psi_n} = \p{\mathcal{A}}{n}_i \mathds{P}_n.
			\end{align}
			If only the $n$th eigenstate is considered, the projection operator $\mathds{P}_n$ can be omitted.

				In other words, the Berry connection of the $n$th eigenstate can be interpreted simply as the projection of $K_i$ onto that state.

			\subsection{Berry Curvature}

				Having established the relationship between the Berry connection and the full Hilbert space bundle connection, it is straightforward to verify the independence of the Berry curvature from the residual gauge freedom under the condition $[\partial_t K_i, H] = 0$.

				Given that the Berry curvature is defined as
			\begin{align}
				\p{\Omega}{n}_{ij} = & \partial_i \p{\mathcal{A}}{n}_j - \partial_j \p{\mathcal{A}}{n}_i\\
				& = \partial_i \Bra{\psi_n} K_j \Ket{\psi_n} - \partial_j \Bra{\psi_n} K_i \Ket{\psi_n}\\
				& = i \Bra{\psi_n}\left[K_i, K_j\right] \Ket{\psi_n},
			\end{align}
			substituting $K_i = t_0 \p{K}{1}_i + \p{K}{0}_i$ and $K_j = t_0 \p{K}{1}_j + \p{K}{0}_j$ yields
			\begin{align}
				\p{\Omega}{n}_{ij} & = i \Bra{\psi_n} \left[t_0 \p{K}{1}_i + \p{K}{0}_i, t_0 \p{K}{1}_j + \p{K}{0}_j\right] \Ket{\psi_n}\\
				& = i \Bra{\psi_n} \left[\p{K}{0}_i, \p{K}{0}_j\right] \Ket{\psi_n}.
			\end{align}
			The final equality holds because $\Ket{\psi_n}$ and $\Bra{\psi_n}$ are simultaneous right- and left-eigenstates of $\p{K}{1}_i$ and $\p{K}{1}_j$, as enforced by the commutation relations $\left[\p{K}{1}_i, H\right] = 0 = \left[\p{K}{1}_j, H\right]$. As promised in the previous subsection, the time-dependent terms drop out of the Berry curvature expression.

				Furthermore, the term $\Delta K_i$ in the gauge transformation
			\begin{align}
				K_i \rightarrow K'_i = K_i + \Delta K_i,
			\end{align}
			does not contribute to the Berry curvature because $\left[\Delta K_i, H\right] = 0$. This confirms that the choice of the generators $K_i$ under the gauge condition $[\partial_t K_i, H] = 0$ does not affect the Berry curvature.

				In summary, the Berry connection $\p{\mathcal{A}}{n}_i$ is, in general, a projection of $K_i$ onto the $n$th eigenstate $\ket{\Phi_n}$. However, Eq.~\eqref{App:BerryConnection} assumes the eigenstates are functions of the Hamiltonian. Nevertheless, when considering state evolution, an eigenstate may evolve into a different eigenstate if an EP exists, thereby requiring additional treatments~\cite{Tang2022, Guo2023, MehriDehnavi2008}. Consequently, it is more natural to study the evolution of general states in the full Hilbert space rather than restricting the analysis to Hamiltonian eigenstates; providing a unified framework for such general state evolutions is precisely one of the purposes of this study.

		\section{Gauge Degrees of Freedom\label{Appendix:GaugeDOF}}

			In this appendix, we review the gauge symmetry associated with the choice of the parameter evolution generator $K$, following Refs.~\cite{Ju2019, Ju2024}.

			It is well-established that quantum systems are described by the Schr\"{o}dinger equation. Specifically, the evolution of the quantum states is governed by
		\begin{align}
			\nabla_t \ket{\Psi} = \left(\partial_t + i H\right) \ket{\Psi} = 0.
		\end{align}
		Therefore, the main assumption is that the Hamiltonian is an invariant quantity (i.e., we do not consider picture changes~\cite{Ju2025} here).

			\subsection{Gauge Transformation\label{Appendix:GrageTranformation}}

				Nevertheless, even with the same Hamiltonian, the representation of the quantum state is non-unique.

				To be more specific, given two arbitrary states $\ket{\Psi}_{\bf 1}$ and $\ket{\Psi}_{\bf 2}$ (where the boldface subscripts {\bf 1} and {\bf 2} denote two different ``choices of basis/gauges'' to be clarified below), both satisfy the same Schr\"{o}dinger equation provided that
				\begin{align}
					\ket{\Psi}_{\bf 2} = T_n^{-1} \ket{\Psi}_{\bf 1}, \label{App:StateTransformation}
				\end{align}
				where $T_n = T_n(t, q^1, \dots, q^n)$ is an invertible matrix depending on time $t$ and $n$ parameters $\{q^1, \dots, q^n\}$, with the subscript $n$ denoting the number of parameters, satisfying
				\begin{align}
					\partial_t T_n + i [H, T_n] = 0. \label{App:MasterT}
				\end{align}

				To see this, let $\ket{\Psi}_{\bf 1}$ be a physical state that satisfies the Schr\"{o}dinger equation
				\begin{align}
					i \partial_t \ket{\Psi}_{\bf 1} = H \ket{\Psi}_{\bf 1}.
				\end{align}
				It follows that $\ket{\Psi}_{\bf 2}$ also satifies the same Schr\"{o}dinger equation:
				\begin{align}
					i \partial_t \ket{\Psi}_{\bf 2} & = i \partial_t \left(T_n^{-1} \ket{\Psi}_{\bf 1}\right)\\
					& = i \left(\partial_t T_n^{-1}\right) \ket{\Psi}_{\bf 1} + i T_n^{-1} \partial_t \ket{\Psi}_{\bf 1}\\
					& = - i T_n^{-1} \left(\partial_t T_n\right) T_n^{-1} \ket{\Psi}_{\bf 1} + T_n^{-1} H \ket{\Psi}_{\bf 1}\\
					& = - T_n^{-1} [H, T_n] T_n^{-1} \ket{\Psi}_{\bf 1} + T_n^{-1} H \ket{\Psi}_{\bf 1}\\
					& = H T_n^{-1} \ket{\Psi}_{\bf 1} = H \ket{\Psi}_{\bf 2}.
				\end{align}

				Indeed, $\ket{\Psi}_{\bf 1}$ and $\ket{\Psi}_{\bf 2}$ are physically equivalent if we apply the transformations
				\begin{align}
					\left\{\begin{array}{rcl}
						\ket{\Psi}_{\bf 1} & \rightarrow & \ket{\Psi}_{\bf 2} = T_n^{-1} \ket{\Psi}_{\bf 1}\\
						G_{\bf 1} & \rightarrow & G_{\bf 2} = T_n^\dagger G_{\bf 1} T_n\\
						\mathcal{O}_{\bf 1} & \rightarrow & \mathcal{O}_{\bf 2} = T_n^{-1} \mathcal{O}_{\bf 1} T_n
					\end{array}\right.,
				\end{align}
				provided $T_n$ satisfies Eq.~\eqref{App:MasterT}, where $\mathcal{O}_{\bf 1}$ is an arbitrary operator in basis 1 and $\mathcal{O}_{\bf 2}$ is the corresponding operator in basis 2. Consequently, the matrix $T_n$ allows us to map states and operators directly from basis 1 to basis 2.

			\subsection{Gauge Transformations and Evolution Generators}

				This subsection investigates the transformation properties of the evolution generators under the action of $T_n$ and vice versa.

				For simplicity, we begin the discussion with just one parameter in addition to time $t$. We now show that the transformation $T_1$ maps one parameter evolution generator to another.

				Let $q$ be a parameter in the Hamiltonian $H$. In Basis 1, the state evolution along the $q$-direction is governed by
				\begin{align}
					\partial_q \ket{\Psi}_{\bf 1} = -i K_{\bf 1} \ket{\Psi}_{\bf 1},
				\end{align}
				where, according to Eq.~\eqref{KEq}, the generator $K_{\bf 1}$ satisfies
				\begin{align}
					\partial_t K_{\bf 1} = i [K_{\bf 1}, H] + \partial_q H.
				\end{align}
				Applying the basis transformation shown in Eq.~\eqref{App:StateTransformation}, the state evolution in Basis 2 becomes
				\begin{align}
					\partial_q \ket{\Psi}_{\bf 2} & = \partial_q \left(T_1^{-1} \ket{\Psi}_{\bf 1}\right)\\
					& = - T_1^{-1} \left(\partial_q T_1\right) T_1^{-1} \ket{\Psi}_{\bf 1} + T_1^{-1} \partial_q \ket{\Psi}_{\bf 1}\\
					& = - T_1^{-1} \left(\partial_q T_1\right) T_1^{-1} \ket{\Psi}_{\bf 1} - i T_1^{-1} K_{\bf 1} \ket{\Psi}_{\bf 1}\\
					& = - T_1^{-1} \left(\partial_q T_1\right) \ket{\Psi}_{\bf 2} - i T_1^{-1} K_{\bf 1} T_1 \ket{\Psi}_{\bf 2}\\
					& = - i K_{\bf 2} \ket{\Psi}_{\bf 2}.
				\end{align}
				Here, the evolution generator $K_{\bf 2}$ in Basis 2 is related to $K_{\bf 1}$ by
				\begin{align}
					K_{\bf 2} = T_1^{-1} K_{\bf 1} T_1 - i T_1^{-1} \left(\partial_q T_1\right). \label{App:KTransform}
				\end{align}

				It is straightforward to show that
				\begin{align}
					\partial_t K_{\bf 2} & = \partial_t \left[T_1^{-1} K_{\bf 1} T_1 - i T_1^{-1} \left(\partial_q T_1\right)\right]\\
					\begin{split}
						& = \left(\partial_t T_1^{-1}\right) K_{\bf 1} T_1 + T_1^{-1} \left(\partial_t K_{\bf 1}\right) T_1 + T_1^{-1} K_{\bf 1} \left(\partial_t T_1\right) - i \left(\partial_t T_1^{-1}\right) \left(\partial_q T_1\right)\\
						& \quad - i T_1^{-1} \partial_t \partial_q T_1 \label{App:DtDqT}
					\end{split}\\
					\begin{split}
						& = i T_1^{-1} [H, T_1] T_1^{-1} K_{\bf 1} T_1 + T_1^{-1} \left(i [K_{\bf 1}, H] + \partial_q H\right) T_1 -i T_1^{-1} K_{\bf 1} [H, T_1]\\
						& \quad + T_1^{-1} [H, T_1] T_1^{-1} \partial_q T_1 - i T_1^{-1} \partial_q \partial_t T_1 \label{App:DqDtT}
					\end{split}\\
					\begin{split}
						& = - i H T_1^{-1} K_{\bf 1} T_1 + T_1^{-1} \left(\partial_q H\right) T_1 + i T_1^{-1} K_{\bf 1} T_1 H + T_1^{-1} H \partial_q T_1 - H T_1^{-1} \partial_q T_1\\
						&\quad - T_1^{-1} [\partial_q H, T_1] - T_1^{-1} [H, \partial_q T_1]
					\end{split}\\
					& = - i H \left(T_1^{-1} K_{\bf 1} T_1 - i T_1^{-1} \partial_q T_1\right) + \partial_q H + i \left(T_1^{-1} K_{\bf 1} T_1 - i T_1^{-1} \partial_q T_1\right) H\\
					& = i [K_{\bf 2}, H] + \partial_q H,
				\end{align}
				where we exchanged the order of differentiation, using $\partial_t \partial_q T_1 = \partial_q \partial_t T_1$ in passing from \eqref{App:DtDqT} to \eqref{App:DqDtT}. In other words, the validity of Eq.~\eqref{KEq} for the evolution generator $K_{\bf 1}$ guarantees its validity for $K_{\bf 2}$, i.e.,
				\begin{align}
					\partial_t K_{\bf 1} = i [K_{\bf 1}, H] + \partial_q H \quad \Leftrightarrow \quad \partial_t K_{\bf 2} = i [K_{\bf 2}, H] + \partial_q H.
				\end{align}

				Conversely, for any pair of evolution generators $K_{\bf 1}$ and $K_{\bf 2}$ that satisfy Eq.~\eqref{KEq}, one can construct the corresponding gauge transformation matrix $T_1$ via Eqs.~\eqref{App:MasterT} and \eqref{App:KTransform}, namely
				\begin{align}
					\left\{\begin{array}{l}
						\partial_t T_1 = i [T_1, H]\\
						\partial_q T_1 = i T_1 K_{\bf 2} - i K_{\bf 1} T_1
					\end{array}\right.. \label{App:DtAndDq}
				\end{align}

				In fact, the reverse direction is nontrivial, as it hinges on the commutativity of partial derivatives, $\partial_q \partial_t T_1 = \partial_t \partial_q T_1$. Nevertheless, applying \eqref{App:DtAndDq} yields
				\begin{align}
					\partial_q \partial_t T_1 - \partial_t \partial_q T_1 & = i \partial_q [T_1, H] - i \partial_t \left(T_1 K_{\bf 2} - K_{\bf 1} T_1\right)\\
					\begin{split}
						& = i [\partial_q T_1, H] + i [T_1, \partial_q H] - i \left(\partial_t T_1\right) K_{\bf 2} - i T_1 \left(\partial_t K_{\bf 2}\right) + i \left(\partial_t K_{\bf 1}\right) T_1\\
						& \quad + i K_{\bf 1} \left(\partial_t T_1\right) 
					\end{split}\\
					\begin{split}
						& = - [T_1 K_{\bf 2} - K_{\bf 1} T_1, H] + i [T_1, \partial_q H] + [T_1, H] K_{\bf 2}\\
						& \quad - i T_1 \left(\partial_t K_{\bf 2}\right) + i \left(\partial_t K_{\bf 1}\right) T_1 - K_{\bf 1} [T_1, H]
					\end{split}\\
					& = i \left(\partial_t K_{\bf 1} - i [K_{\bf 1}, H] - \partial_q H\right) T_1 - i T_1 \left(\partial_t K_{\bf 2} - i [K_{\bf 2}, H] - \partial_q H\right).
				\end{align}
				Since $K_{\bf 1}$ and $K_{\bf 2}$ both satisfy Eq.~\eqref{KEq}, namely,
				\begin{align}
					\partial_t K_{\bf 1} = i [K_{\bf 1}, H] + \partial_q H \quad \text{and} \quad \partial_t K_{\bf 2} = i [K_{\bf 2}, H] + \partial_q H,
				\end{align}
				the partial derivatives commute for $T_1$, i.e.,
				\begin{align}
					\partial_q \partial_t T_1 = \partial_t \partial_q T_1.
				\end{align}

				Therefore, given any two arbitrary evolution generators $K_{\bf 1}$ and $K_{\bf 2}$, they are related by a gauge transformation.

				\noindent\emph{Extension to $n$ parameters.} The derivation for the multi-parameter case requires no new machinery and simply repeats the two-parameter calculations. Applying the same method, we can show that two sets of evolution generators, $\{K_i\}$ and $\{K'_i\}$, are related by a single gauge transformation matrix $T_n$ satisfying
				\begin{align}
					\left\{\begin{array}{l}
						\partial_0 T_n = i [T_n, H]\\
						\partial_i T_n = i T_n K'_i - i K_i T_n
					\end{array}\right..
				\end{align}
				(Recall that $\partial_0 = \partial/\partial t$ and $\partial_i = \partial/\partial q^i$.)

				Similar to the one-parameter case, different choices of the evolution generators $\{K_i\}$ merely correspond to different gauges.

		\section{Gauge Invariance of the Holonomy Class \label{Appendix:HolonomyClass}}

			Following the discussion in Appendix~\ref{Appendix:GaugeDOF}, the gauge covariance (or gauge equivalence) of the holonomy group is straightforward to establish.

			To see this equivalence, let $\mathcal{C}_a$ and $\mathcal{C}_b$ be two loops parameterized by $s \in [0,1]$ based at the same point $\tilde{q}$ in the space-time parameter space $\mathds{R} \times \mathcal{M}^n$, such that $\mathcal{C}_a(0) = \mathcal{C}_b(0) = \tilde{q}$. The product loop $\mathcal{C}_c = \mathcal{C}_b * \mathcal{C}_a$ is defined by
			\begin{align}
				\mathcal{C}_c (s) = \mathcal{C}_b * \mathcal{C}_a (s) = \left\{\begin{array}{ll}
					\mathcal{C}_a(2s), & 0 \leq s \leq \frac{1}{2}\\
					\mathcal{C}_a(2s - 1), & \frac{1}{2} < s \leq 1
				\end{array}\right.,
			\end{align}
			which satisfies $\mathcal{C}_c(0) = \mathcal{C}_c(1) = \tilde{q}$.

			Under Gauge 1 (denoted by subscript $\bf 1$), the states evolving along these loops satisfy
			\begin{align}
				\begin{split}
					& \ket{\Psi[\mathcal{C}_a](s = 1)}_{\bf 1} = U_a \ket{\Psi[\mathcal{C}_a](s = 0)}_{\bf 1},\\
					& \ket{\Psi[\mathcal{C}_a](s = 1)}_{\bf 1} = U_b \ket{\Psi[\mathcal{C}_b](s = 0)}_{\bf 1},\\
					& \ket{\Psi[\mathcal{C}_c](s = 1)}_{\bf 1} = U_c \ket{\Psi[\mathcal{C}_c](s = 0)}_{\bf 1} = U_b \ket{\Psi[\mathcal{C}_b](s = 0)}_{\bf 1}\\
					&\hspace{2.35cm} = U_b U_a \ket{\Psi[\mathcal{C}_a](s = 0)}_{\bf 1} = U_b U_a \ket{\Psi[\mathcal{C}_c](s = 0)}_{\bf 1},
				\end{split}\label{App:Holonomy1}
			\end{align}
			where $U_a, U_b,$ and $U_c$ are the corresponding holonomies.

			Similarly, under Gauge 2, the holonomies $U'_a, U'_b,$ and $U'_c$ associated with the same loops based at $\tilde{q}$ are defined by
			\begin{align}
				\begin{split}
					& \ket{\Psi[\mathcal{C}_a](s = 1)}_{\bf 2} = U'_a \ket{\Psi[\mathcal{C}_a](s = 0)}_{\bf 2},\\
					& \ket{\Psi[\mathcal{C}_a](s = 1)}_{\bf 2} = U'_b \ket{\Psi[\mathcal{C}_b](s = 0)}_{\bf 2},\\
					& \ket{\Psi[\mathcal{C}_c](s = 1)}_{\bf 2} = U'_c \ket{\Psi[\mathcal{C}_c](s = 0)}_{\bf 2} = U'_b U'_a \ket{\Psi[\mathcal{C}_c](s = 0)}_{\bf 2}.
				\end{split}\label{App:Holonomy2}
			\end{align}

			As shown in Appendix~\ref{Appendix:GrageTranformation}, quantum states in Gauge 1 and Gauge 2 are related by
			\begin{align}
				\ket{\Psi}_{\bf 2} = T_n^{-1} \ket{\Psi}_{\bf 1}, \label{App:StateTransform2}
			\end{align}
			where $T_n = T_n(t, q^1, \dots, q^n)$ is the gauge transformation operator. Because $\mathcal{C}_a(0) = \mathcal{C}_a(1) = \mathcal{C}_b(0) = \mathcal{C}_b(1) = \tilde{q}$ and $T_n$ is a single-valued function of position, $T_n$ takes the same value at all endpoint base points. For brevity, we define $T \equiv T_n\vert{}_{\tilde{q}}$.

			Combining Eqs.~\eqref{App:Holonomy1}, \eqref{App:Holonomy2}, and \eqref{App:StateTransform2}, the holonomies in the two gauges are related by a similarity transformation, namely,
			\begin{align}
				U'_a = T^{-1} U_a T, \quad U'_b = T^{-1} U_b T, \quad \text{and} \quad U'_c = T^{-1} U_c T.
			\end{align}
	
				Consequently, for any arbitrary loops $\mathcal{C}_a$ and $\mathcal{C}_b$,
			\begin{align}
				U'_c = U'_b U'_a = T^{-1} U_b T T^{-1} U_a T = T^{-1} U_b U_a T = T^{-1} U_c T
			\end{align}
			demonstrating that the holonomies are related by a global similarity transformation. In other words, the holonomy group elements under two different gauges are in one-to-one correspondence, fully preserving the group structure under gauge transformations.

			As a consequence, the holonomy groups under different gauge choices are isomorphic. Hence, the holonomy class is invariant under gauge transformations.

		\section{Evolution Operator for Nontrivial Time Loop \label{Appendix:EvolutionOperators}}

			This Appendix provides the derivation of the solution $U[\widetilde{\gamma}_-]$ for Eq.~\eqref{Utildegamma} with the initial condition $U[\widetilde{\gamma}_-](0; 0) = \mathds{1}$, keeping these detailed calculations out of the main text.

			To significantly simplify the calculations, we switch to the Pauli matrix basis. In this representation, the parameter evolution generators become
			\begin{align}
				H & = (1 + i y) \sigma_x - i x \sigma_z = \bm{R} \cdot \bm{\sigma},\\
				K_x & = \frac{-(1 + i y) x t}{(1 + i y)^2 - x^2} \sigma_x + \frac{i (1 + i y)}{(1 + i y)^2 - x^2} \sigma_y + \frac{i x^2 t}{(1 + i y)^2 - x^2} \sigma_z\\
				& = \left[t \frac{\bm{R} \cdot \left(\partial_x \bm{R}\right)}{R^2} \bm{R} + \frac{\bm{R} \times \left(\partial_x \bm{R}\right)}{2 R^2}\right] \cdot \bm{\sigma},\\
				K_y & = \frac{i(1 + i y)^2 t}{(1 + i y)^2 - x^2} \sigma_x + \frac{x / 2}{(1 + i y)^2 - x^2} \sigma_y + \frac{(1 + i y) x t}{(1 + i y)^2 - x^2} \sigma_z\\
				& = \left[t \frac{\bm{R} \cdot \left(\partial_y \bm{R}\right)}{R^2} \bm{R} + \frac{\bm{R} \times \left(\partial_y \bm{R}\right)}{2 R^2}\right] \cdot \bm{\sigma},
			\end{align}
			where $\bm{R} = (1 + i y, \, 0, \, - i x)$, $R^2 = \bm{R} \cdot \bm{R} = (1 + i y)^2 - x^2$, and $\bm{\sigma} = (\sigma_x, \sigma_y, \sigma_z)$, with $\sigma_i$ being the standard Pauli matrices:
			\begin{align}
				\sigma_x = \begin{pmatrix}
					0 & 1\\
					1 & 0
				\end{pmatrix}, \quad \sigma_y = \begin{pmatrix}
					0 & - i\\
					i & 0
				\end{pmatrix}, \quad \text{and} \quad \sigma_z = \begin{pmatrix}
					1 & 0\\
					0 & -1
				\end{pmatrix}.
			\end{align}
			Boldface font denotes vectors in this representation to avoid confusion with the $(t, x, y)$-vector in Section~\ref{Sec:Holonomy}.

			From Eq.~\eqref{Utildegamma}, the governing equation for $U[\widetilde{\gamma}_-]$ is given by
			\begin{align}
				\frac{d}{d \theta} U[\widetilde{\gamma}_-](\theta; 0) & = - i \left.\left(T' H + \frac{d x}{d \theta} K_x + \frac{d y}{d \theta} K_y \right)\right|_{\widetilde{\gamma}_-} U[\widetilde{\gamma}_-](\theta; 0). \label{App:NontrivialDEq}
			\end{align}

			To solve Eq.~\eqref{App:NontrivialDEq}, we employ the following matrix identities:
			\begin{align}
				& e^{-i \bm{r} \cdot \bm{\sigma}} = \cos\left(r\right) \mathds{1} - i \frac{\sin\left(r\right)}{r} \bm{r} \cdot \bm{\sigma},\\
				& \left(\bm{r}_1 \cdot \bm{\sigma}\right) \left(\bm{r}_2 \cdot \bm{\sigma}\right) = \bm{r}_1 \cdot \bm{r}_2 \mathds{1} + i \left(\bm{r}_1 \times \bm{r}_2\right) \cdot \bm{\sigma},
			\end{align}
			where $r = (\bm{r} \cdot \bm{r})^{1/2}$. Using these relations, direct calculations yield
			\begin{align}
				\begin{split}
					& e^{i \bm{F} \cdot \bm{\sigma}} \frac{d}{d \theta} e^{-i \bm{F} \cdot \bm{\sigma}}\\
					& \quad = - i \left[\left(\frac{d}{d \theta} F \right)\frac{F - \sin F \cos F}{F^2} \bm{F} + \frac{\sin F \cos F}{F} \left(\frac{d}{d \theta} \bm{F}\right) - \left(\frac{\sin F}{F}\right)^2 \bm{F} \times \left(\frac{d}{d \theta} \bm{F}\right)\right] \cdot \bm{\sigma},
				\end{split} \label{App:MatrixIdentity1}
			\end{align}
			and
			\begin{align}
				\begin{split}
					& e^{i \bm{F} \cdot \bm{\sigma}} \left(\bm{Y} \cdot \bm{\sigma}\right) e^{-i \bm{F} \cdot \bm{\sigma}}\\
					& \quad = \left[(\cos F)^2 - (\sin F)^2] \bm{Y} + \frac{(\sin F)^2}{F^2} \left(\bm{F} \cdot \bm{Y}\right) \bm{F} + \frac{\sin F \cos F}{F} \left(\bm{Y} \times \bm{F}\right)\right] \cdot \bm{\sigma},
				\end{split} \label{App:MatrixIdentity2}
			\end{align}
			where $F = \left(\bm{F} \cdot \bm{F}\right)^{1/2}$.

			We then make the ansatz
			\begin{align}
				U[\widetilde{\gamma}_-](\theta; 0) & = e^{- i T(\theta) \bm{R} (\theta) \cdot \bm{\sigma}} \widetilde{U}(\theta) = \left[\cos \left(T(\theta) R (\theta)\right) \mathds{1} - i \frac{\sin \left(T(\theta) R (\theta)\right)}{R (\theta)} \bm{R} (\theta) \cdot \bm{\sigma}\right] \widetilde{U}(\theta),
			\end{align}
			where $\bm{R} = (1 + i y (\theta), 0, - i x (\theta))$ depends implicitly on $\theta$. Substituting this ansatz into Eq.~\eqref{App:NontrivialDEq} without specifying the explicit forms of $x(\theta)$ and $y(\theta)$, and multiplying by $\exp\left(i T(\theta) \bm{R}_- (\theta) \cdot \bm{\sigma} \right)$ from the left, we obtain
			\begin{align}
				\begin{split}
					& \underbrace{e^{i T(\theta) \bm{R}_- (\theta) \cdot \bm{\sigma}} \left(\frac{d}{d \theta} e^{- i T(\theta) \bm{R}_- (\theta) \cdot \bm{\sigma}}\right)}_{\text{\ding{172}}} \widetilde{U}(\theta) + \frac{d}{d \theta} \widetilde{U}(\theta)\\
					& \quad = \underbrace{- i e^{i T(\theta) \bm{R}_- (\theta) \cdot \bm{\sigma}} \left.\left(\frac{d t}{d \theta} H + \frac{d x}{d \theta} K_x + \frac{d y}{d \theta} K_y \right)\right|_{\widetilde{\gamma}_-} e^{i T(\theta) \bm{R}_- (\theta) \cdot \bm{\sigma}}}_{\text{\ding{173}}} \widetilde{U}(\theta).
				\end{split} \label{App:Ansatz}
			\end{align}
			These numbered terms are highlighted for reference and will be utilized in the discussion below.

			Noting that
			\begin{align}
				\begin{split}
					& \frac{d t}{d \theta} H + \frac{d x}{d \theta} K_x + \frac{d y}{d \theta} K_y\\
					& \quad = \left\{\frac{d t}{d \theta} \bm{R} + \frac{d x}{d \theta} \left[t \frac{\bm{R} \cdot \left(\partial_x \bm{R}\right)}{R^2} \bm{R} + \frac{\bm{R} \times \left(\partial_x \bm{R}\right)}{2 R^2}\right] + \frac{d y}{d \theta} \left[t \frac{\bm{R} \cdot \left(\partial_y \bm{R}\right)}{R^2} \bm{R} + \frac{\bm{R} \times \left(\partial_y \bm{R}\right)}{2 R^2}\right]\right\} \cdot \bm{\sigma}
				\end{split}
			\end{align}
			and that $\bm{R}$ depends solely on $x$ and $y$, this combination simplifies to
			\begin{align}
				\frac{d t}{d \theta} H + \frac{d x}{d \theta} K_x + \frac{d y}{d \theta} K_y = \left[\frac{d t}{d \theta} \bm{R} + \left(t \frac{\bm{R} \cdot \bm{R}'}{R^2} \bm{R} + \frac{\bm{R} \times \bm{R}'}{2 R^2}\right)\right] \cdot \bm{\sigma},
			\end{align}
			where $\bm{R}' = \dfrac{d}{d \theta} \bm{R}$.

			Applying the identities in Eqs.~\eqref{App:MatrixIdentity1} and \eqref{App:MatrixIdentity2}, a straightforward calculation shows that
			\begin{align}
				\begin{split}
					& - i e^{i T(\theta) \bm{R}_- (\theta) \cdot \bm{\sigma}} \left[\underbrace{\left.\left(T' H + \frac{d x}{d \theta} K_x + \frac{d y}{d \theta} K_y \right)\right|_{\widetilde{\gamma}_-} e^{- i T(\theta) \bm{R}_- (\theta) \cdot \bm{\sigma}}}_{\text{\ding{173}}} - \underbrace{\left(\frac{d}{d \theta} e^{- i T(\theta) \bm{R}_- (\theta) \cdot \bm{\sigma}}\right)}_{\text{\ding{172}}}\right]\\
					& \quad = \frac{- (1 - \rho e^{- i \theta})}{2 (2 - \rho e^{- i \theta})} \sigma_y,
				\end{split}
			\end{align}
			where $\bm{R}_-(\theta) = (1 + i \rho \sin \theta, 0, i (1 - \rho \cos \theta))$, $R_- (\theta) = \left[\bm{R}_- (\theta) \cdot \bm{R}_- (\theta)\right]^{1 / 2} = e^{i \theta / 2} \rho^{1 / 2} \left(2 - \rho e^{- i \theta}\right)^{1 / 2}$. Note that these expressions match the respective components designated as \ding{172} and \ding{173} in Eq.~\eqref{App:Ansatz}.

			Consequently, Eq.~\eqref{App:Ansatz} reduces to the differential equation
			\begin{align}
				\Rightarrow & \frac{d}{d \theta} \widetilde{U}(\theta) = \frac{- (1 - \rho e^{- i \theta})}{2 (2 - \rho e^{- i \theta})} \sigma_y \widetilde{U}(\theta),\label{App:TildeUDEq}
			\end{align}
			which matches the differential equation given in Eq.~\eqref{UGammaMinus}.

			The solution to Eq.~\eqref{App:TildeUDEq} is thus given by
			\begin{align}
				\widetilde{U}(\theta) = U[\gamma_-](\theta; 0) = \frac{1}{2 \lambda_-(\theta)} \begin{pmatrix}
					1 + \lambda_-(\theta)^2 & - i \left(1 - \lambda_-(\theta)^2\right)\\
					i \left(1 - \lambda_-(\theta)^2\right) & 1 + \lambda_-(\theta)^2
				\end{pmatrix},
			\end{align}
			where both $U[\gamma_-] (\theta; 0)$ and $\lambda_-(\theta)$ are defined as in Sec.~\ref{Sec:EnclosingEP}.

			Substituting this back into the ansatz yields the final solution for Eq.~\eqref{App:NontrivialDEq} renders
			\begin{align}
				U[\widetilde{\gamma}_-](\theta; 0) & = e^{i T(\theta) \bm{R}_- (\theta) \cdot \bm{\sigma}} \widetilde{U}(\theta)\\
				\begin{split}
					& = \frac{1}{2 \lambda_-(\theta)} \left[\cos \left(T(\theta) R_- (\theta)\right) \mathds{1} - i \frac{\sin \left(T(\theta) R_- (\theta)\right)}{R_- (\theta)} \bm{R}_- (\theta) \cdot \bm{\sigma}\right]\big[\left(1 + \lambda_-(\theta)^2\right) \mathds{1}\\
					& \quad + \left(1 - \lambda_-(\theta)^2\right) \sigma_y\big]
				\end{split}\\
				\begin{split}
					& = \frac{(1 + \lambda_- (\theta)^2)}{2 \lambda_- (\theta)} \cos \left(T (\theta) R_- (\theta)\right) \mathds{1} - \dfrac{i \left(2 - \rho e^{- i \theta} + \lambda_- (\theta)^2 \rho e^{i \theta}\right)}{2 R_- (\theta) \lambda_- (\theta)} \sin \left(T (\theta) R_- (\theta)\right) \sigma_x\\
					& \quad + \frac{\left(1 - \lambda_- (\theta)^2\right)}{2 \lambda_- (\theta)} \cos \left(T (\theta) R_- (\theta)\right) \sigma_y + \dfrac{\left(2 - \rho e^{- i \theta} - \lambda_- (\theta)^2 \rho e^{i \theta}\right)}{2 R_- (\theta) \lambda_- (\theta)} \sin \left(T (\theta) R_- (\theta)\right) \sigma_z.
				\end{split}
			\end{align}
	
			This confirms that the evolution along paths $\widetilde{\gamma}_1$ and $\widetilde{\gamma}_2$ (see Fig.~\ref{Fig:Equivalence}) yields identical results.
	
	\bibliography{References}
\end{document}